\documentclass[12pt,english]{article}
\usepackage[latin9]{inputenc}
\usepackage{geometry}
\usepackage{float}
\usepackage{booktabs}
\usepackage{footnote}
\usepackage{amsmath}
\usepackage{amsthm}
\usepackage{amssymb}
\usepackage{graphicx}
\usepackage{setspace}
\usepackage[authoryear]{natbib}

\makeatletter

\makesavenoteenv{tabular}

\providecommand{\tabularnewline}{\\}

\numberwithin{equation}{section}
\theoremstyle{plain}
\newtheorem{thm}{\protect\theoremname}[section]
\theoremstyle{definition}
\newtheorem{rem}{\protect\remarkname}[section]
\theoremstyle{plain}
\newtheorem{cor}{\protect\corollaryname}[section]

\usepackage{setspace}
\usepackage{sectsty}
\usepackage{datetime}
\usepackage{times}
\usepackage{amsthm}
\usepackage{amsfonts}
\usepackage{amsmath}
\usepackage{amssymb}

\sectionfont{\noindent\normalfont\large\bf}
\subsectionfont{\noindent\normalfont\normalsize\bf}
\subsubsectionfont{\noindent\normalfont\bf}

\ifx\pdfoutput\@undefined\usepackage[usenames,dvips]{color}
\else\usepackage[usenames,dvipsnames]{color}
\IfFileExists{pdfcolmk.sty}{\usepackage{pdfcolmk}}{} 
\fi

\usepackage[colorlinks,breaklinks,   
     bookmarks=false,
     pdfstartview=Fit,  
     pdfview=Fit,       
     linkcolor=Black,
     urlcolor=Black,
     citecolor=Black,
     hyperfootnotes=false
         ]{hyperref}
\usepackage[figure,table]{hypcap} 
\interfootnotelinepenalty=\@M
\allowdisplaybreaks
\AtBeginDocument{
  
}

\makeatother

\usepackage{babel}
\providecommand{\corollaryname}{Corollary}
\providecommand{\remarkname}{Remark}
\providecommand{\theoremname}{Theorem}

\begin{document}
\title{\singlespacing{}\noindent \textbf{\Large{}Uncertainty in the Hot Hand Fallacy: }\\
\textbf{\Large{} Detecting Streaky Alternatives to Random Bernoulli
Sequences}}
\author{\noindent %
\begin{tabular}{ccc}
David M. Ritzwoller & $ $ & Joseph P. Romano\textbf{}\thanks{E-mail:\ ritzwoll@stanford.edu, romano@stanford.edu. DR acknowledges
funding from the Stanford Institute for Economic Policy Research and
the National Science Foundation under the Graduate Research Fellowship
Program. JR acknowledges funding from the National Science Foundation
(MMS-1949845). We thank Tom DiCiccio, Maya Durvasula, Matthew Gentzkow,
Tom Gilovich, Zong Huang, Victoria de Quadros, Joshua Miller, Linda
Ouyang, Adam Sanjurjo, Azeem Shaikh, Jesse Shapiro, Hal Stern, Marius
Tirlea, Shun Yang, Molly Wharton, Michael Wolf, and seminar audiences
at Stanford University and the California Econometrics Conference
for helpful comments and conversations.}\tabularnewline
{\small{}Stanford University} & $ $ & {\small{}Stanford University}\tabularnewline
\end{tabular}\textbf{}\\
}
\date{\large{\monthname\ \number\day, \number\year}}

\maketitle
\vspace{-0.3in}
\begin{abstract}
We study a class of permutation tests of the randomness of a collection
of Bernoulli sequences and their application to analyses of the human
tendency to perceive streaks of consecutive successes as overly representative
of positive dependence---the hot hand fallacy. In particular, we
study permutation tests of the null hypothesis of randomness (i.e.,
that trials are i.i.d.) based on test statistics that compare the
proportion of successes that directly follow $k$ consecutive successes
with either the overall proportion of successes or the proportion
of successes that directly follow $k$ consecutive failures. We characterize
the asymptotic distributions of these test statistics and their permutation
distributions under randomness, under a set of general stationary
processes, and under a class of Markov chain alternatives, which allow
us to derive their local asymptotic power. The results are applied
to evaluate the empirical support for the hot hand fallacy provided
by four controlled basketball shooting experiments. We establish that
substantially larger data sets are required to derive an informative
measurement of the deviation from randomness in basketball shooting.
In one experiment, for which we were able to obtain data, multiple
testing procedures reveal that one shooter exhibits a shooting pattern
significantly inconsistent with randomness -- supplying strong evidence
that basketball shooting is not random for all shooters all of the
time. However, we find that the evidence against randomness in this
experiment is limited to this shooter. Our results provide a mathematical
and statistical foundation for the design and validation of experiments
that directly compare deviations from randomness with human beliefs
about deviations from randomness, and thereby constitute a direct
test of the hot hand fallacy.

\smallskip{}


\smallskip{}


\bigskip{}

\setcounter{page}{0}
\thispagestyle{empty}
\end{abstract}
\pagebreak{}

\begin{spacing}{1.4}

\section{Introduction}

Suppose that we observe $s$ Bernoulli sequences of length $n$. We
are interested in testing the null hypothesis that these sequences
are independent and identically distributed (i.i.d.)\ against alternatives
in which the probability of success following a streak of consecutive
successes is greater than it is either unconditionally or following
a streak of consecutive failures. The interpretation of results of
tests of this form have been pivotal in the development of behavioral
economics.

In an influential paper, \citet{tversky1971belief} hypothesize that
people tend to believe that small samples are overly representative
of the ``essential characteristics'' of the population from which
they are drawn. They describe this phenomenon as ``belief in the
law of small numbers,'' and present several evocative examples in
support of this claim. For instance, they show that academic researchers
tend to substantially underestimate sample sizes necessary to achieve
adequate statistical power against reasonable alternatives in the
design of experiments -- evidently finding samples overly representative
of the populations from which they are drawn. Similarly, in what has
been termed the ``gambler's fallacy,'' when asked to successively
predict outcomes of an i.i.d.\ Bernoulli sequence, experimental subjects
tend to underestimate the probability of streaks of consecutive successes
or failures -- apparently perceiving streaks to be overly representative
of non-randomness. Subsequently, this insight into misperception of
randomness has been formalized \citep{rabin2002inference} and integrated
into standard models in behavioral economics and finance \citep{barberis2003survey,barberis2018psychology}.\footnote{\citet{bar1991perception} review the psychological literature on
models of misperception of randomness, highlighting their implications
for judgment of dependence in random Bernoulli sequences. \citet{benjamin2019errors}
reviews the psychological and behavioral economics literatures on
errors in probabilistic reasoning, surveying the available empirical
support for proposed biases and highlighting areas of economics where
these biases are relevant.}

The ``hot hand fallacy,'' proposed and studied in \citet*{gilovich1985hot},
henceforth GVT, is a behavioral bias attributable to belief in the
law of small numbers. As subsequently formalized in \citet{rabin2010the},
the hot hand fallacy refers to a positive bias in beliefs about the
dependence in a Bernoulli process following the observation of a streak
of consecutive successes. In particular, when faced with a streak
of consecutive successes, believers in the law of small numbers overestimate
the positive dependence in the sequence -- perceiving streaks to
be overly representative of dependence.\footnote{We refer the reader to \citet{rabin2010the} for a precise analysis
of conditions under which belief in the law of small numbers implies
the hot hand and gambler's fallacies.} GVT aim to document the hot fallacy using data from a controlled
basketball shooting experiment and results from a survey on beliefs
in the serial dependence of basketball shooting. They fail to reject
the hypothesis that the sequences of shots they observe are i.i.d.,
but document a widespread belief in the ``hot hand'' -- that basketball
players are more likely to make a shot after one or more successful
shots than after one or more misses. Thus, they conclude that the
belief in the hot hand is a pervasive cognitive illusion or fallacy,
giving provocative evidence in favor of models that incorporate belief
in the law of small numbers. This result became the academic consensus
for the following three decades \citep{kahneman2011thinking,thaler2009nudge},
and provided a central empirical support for economic models in which
agents are overconfident in conclusions drawn from small samples \citep{barberis2003survey}.

The GVT results were challenged by \citet{miller2018surprised}, henceforth
MS, who discovered a significant small-sample bias in plug-in estimates
of the probability of success following streaks of successes or failures.
They argue that when they correct the GVT analysis for this small-sample
bias, they are able to reject the null hypothesis that shots are i.i.d.,
in favor of positive dependence that is consistent with expectations
of streakiness in basketball shooting.\footnote{The MS results received extensive coverage in the popular press, including
expository articles in the New York Times (\citealt{johnson2015gamblers}
and \citealt{appelbaum2015streaks}), the New Yorker \citep{remnick2017bob},
the Wall Street Journal \citep{cohen2015hot}, and on ESPN \citep{habersttroh2017hes},
among other media outlets. MS was the 10th most downloaded paper on
SSRN in 2015. Statistics sourced from http://ssrnblog.com/2015/12/29/ssrn-top-papers-of-2015/,
accessed on July 21st, 2019.} \citet{miller2018momentum} argue that their work ``uncovered critical
flaws ... sufficient to not only invalidate the most compelling evidence
against the hot hand, but even to vindicate the belief in streakiness.''
A more conservative interpretation of their conclusions suggests that
their work creates persisting uncertainty about the empirical support
for textbook theories of misperception of randomness. \citet{benjamin2019errors}
writes that MS ``re-opens--but does not answer--the key question
of whether there is a hot hand\textit{ bias} ... a belief in a stronger
hot hand than there really is.''

The objective of this paper is to clarify and quantify the uncertainty
in the evidence that controlled basketball shooting experiments have
contributed to our understanding of the hot hand fallacy and, by implication,
economic models incorporating belief in the law of small numbers.
Towards this goal, we develop a formal statistical framework for testing
the randomness of a set of Bernoulli sequences and measure the finite-sample
power of these tests with local asymptotic approximations. Equipped
with these theoretical results, we then provide a comprehensive analysis
of the design and interpretation of the outcomes of four controlled
basketball shooting experiments and give recommendations for the design
and methodology of future empirical work.

Section \ref{sec:Posing-the-Problem} develops a formal statistical
framework for assessing the positive serial dependence in basketball
shooting using data from controlled shooting experiments. We emphasize
the distinctions between individual, simultaneous, and joint hypothesis
testing. We specify our null hypothesis -- that observed shooting
outcomes are i.i.d.\ -- and a set of alternative hypotheses in which
the probability of a make following a streak of consecutive makes
is greater than it is either unconditionally or following a streak
of consecutive misses. We denote this class of alternatives as ``streaky'',
as the probability of a streak of makes is larger than it would be
under randomness. These alternatives motivate a set of natural plug-in
test statistics, studied previously in GVT and MS: the observed differences
between the proportions of makes following a streak of consecutive
makes and either the overall proportion of makes or the proportion
of makes following a streak of consecutive misses.

Section \ref{sec:Detecting} develops methods for testing the randomness
of a collection of Bernoulli sequences against streaky alternatives
using these test statistics. We derive the asymptotic distributions
of the test statistics specified in Section \ref{sec:Posing-the-Problem}
under the null hypothesis of randomness and under general stationary
alternatives. We highlight the substantial small-sample biases of
these approximations under the null hypothesis, which were discovered
and studied in MS. This bias motivates the application of permutation
tests, which we show are the only tests with exact type 1 error control.
We conclude the Section by characterizing the asymptotics of the test
statistics' permutation distributions under the null hypothesis and
under general stationary alternatives.

In Section \ref{sec:Power}, these results allow us to derive asymptotic
approximations to the power of the permutation tests developed in
Section \ref{sec:Detecting} against a specific class of Markov chain
streaky alternatives with a local asymptotic approximation. These
results significantly reduce the computational expense of power analyses
in the design of future experiments. Simulation evidence indicates
that our asymptotic power approximations perform remarkably well in
the sample sizes considered in available controlled basketball shooting
experiments.

Despite the long history of the tests that we study, our asymptotic
results are new. Though some of our initial arguments are fairly standard,
deriving the limiting behavior of the permutation distributions proved
challenging, even under the null. The standard approach is to verify
Hoeffding's condition; see Theorem \ref{thm: General D_1}. To do
so, we develop a novel application of the \citet{rinott1994on} central
limit theorem, which is based on Stein's method. Our derivation of
the limiting behavior and local power of the permutation tests under
dependent processes is more complex. We obtain the limiting behavior
of the permutation distribution under deterministic sequences (i.e.,
when the number of successes is fixed) with a novel equicontinuity
argument (Lemma K.1 in the Online Appendix). This result (Lemma K.2
in the Online Appendix) holds without probabilistic qualification,
unlike results obtained from verifying Hoeffding's condition, and
allows us to derive the limiting behavior of the test statistics under
dependent sequences (Theorem \ref{thm: General D_1}).

Having developed a formal statistical framework for testing the randomness
of a collection of Bernoulli sequences, in Section \ref{sec:hot-hand}
we evaluate the implications of the outcomes of four controlled basketball
shooting experiments for the question posed in \citet{benjamin2019errors}:
``whether there is ... a belief in a stronger hot hand than there
really is.'' A conclusive answer to this question requires informative
estimates of the actual deviation from randomness and expectations
of the deviation from randomness in basketball shooting.

First, we analyze the design and results of four controlled basketball
shooting experiments. We find that there is strong evidence that basketball
shooting is not perfectly random for all basketball players all of
the time. In data from the GVT experiment, we find that we are able
to reject i.i.d.\ shooting consistently after accounting for multiplicity
for only one shooter out of twenty-six, identified in the dataset
as ``Shooter 109''. This shooter's shot sequence is remarkably streaky:
he makes 16 shots in a row directly following a period in which he
misses 15 out of 18 shots.\footnote{GVT observe the rejection of the null hypothesis for Shooter 109,
but concede ``we might expect one significant result out of 26 by
chance.'' We show that the rejection of the null hypothesis for Shooter
109 is robust to standard multiple testing corrections. \citet{Wardrop1999statistical}
notes that the $p$-value for standard tests of the randomness of
the shooting sequence for Shooter 109 is extremely small.} However, we argue that the four controlled shooting experiments do
not have adequate power to detect parameterizations of the Markov
chain streaky alternative, studied in Section \ref{sec:Power}, consistent
with the variation in NBA field goal shooting percentages.\footnote{Our results align with the conclusions of \citet{stern1993a}, who
show that tests of the randomness of hitting streaks in baseball applied
in \citet{albright1993statistical} have limited power.} Moreover, evidence against randomness in the GVT experiment appears
to be confined to Shooter 109.\footnote{We are not the first to observe that the GVT data are underpowered
for the Markov chain alternatives. \citet{miller2019fallacy}, \citet{miyoshi2000hot},
and \citet{Wardrop1999statistical} measure the power of individual
tests against specific parameterizations of similar models with simulation.
\citet{korb2003story} and \citet{stone2012measurement} measure power
against particular non-stationary alternatives. We contribute to these
analyses by deriving analytical approximations of the power, studying
a significantly richer set of parameterizations of these models, informing
our choices of alternatives by comparison to NBA shooting percentages,
and explicitly considering simultaneous and joint null hypothesis
tests.}

Second, we assess the available evidence on expectations of streakiness.
We highlight methodological limitations of the surveys of basketball
fans and incentivized experiments presented in GVT and MS. We note
a variety of observational estimates \citep{rao2009experts,bocskocsky2014hot,lantis2019hot}
consistent with large expected deviations from randomness, but find
that all available estimates of beliefs are not directly comparable
to measurements of the serial dependence in basketball shooting.

We conclude that larger data and more structured elicitation of beliefs
are required to resolve the uncertainty in the empirical support for
the hot hand fallacy. We provide a mathematical and statistical foundation
for future work with this objective.

Tests of the randomness of stochastic processes against nonrandom,
persistent, or serially dependent alternatives have been studied extensively
within finance and economics; the framework and methods that we develop
are applicable to these settings. In Online Appendix A, we outline
the application of our methods to two problems in empirical finance:
tests of the weak form efficient market hypothesis \citep{fama1970efficient,malkiel2003efficient}
and tests of persistence in the performance of mutual funds relative
to benchmarks \citep{jensen1968performance,hendricks1993hot,carhart1997persistence}.
In particular, in the spirit of \citet{fama1965behavior}, in Online
Appendix A.1 we implement the individual permutation tests that we
develop in Section \ref{subsec:Permutation-Tests} on two datasets
of stock price sequences. Moreover, the problem of testing for and
estimating state dependence -- the causal effect of an outcome in
the previous period on the current period's outcome -- is widely
studied in microeconomics (see e.g., \citealt{heckman1981heterogeneity,chay1999true,keane1997modeling}).
In Online Appendix B, we show that our methods provide a test for
state dependence under appropriate unconfoundedness type assumptions.\footnote{As they do not play a role in our empirical application, we do not
consider covariates or instruments. The extension of our methods to
account for observed and unobserved heterogeneity across time and
individuals is important for their application to these contexts and
may be fruitful. \citet{torgovitsky2019nonparametric} develops a
partial identification approach to bounding state dependence in these
settings.}

Section 6 concludes. Online Appendices A-J give supplementary results
and discussion that will be introduced at appropriate points throughout
the paper. Proofs of all Theorems presented in the main body of this
paper are given in Online Appendix K.

\section{Posing the Problem\label{sec:Posing-the-Problem}}

Do people overestimate positive serial dependence in basketball shooting?
Three components of this question are often conflated:
\begin{itemize}
\item Is there any positive serial dependence in basketball shooting?
\item If so, how widespread and substantial is it?
\item And finally, do people systematically overestimate this dependence?
\end{itemize}
In this section, we provide a formal framework that will enable us
to develop inferential methods for answering the first two questions.
In Section \ref{sec:hot-hand}, we provide a discussion of methods
for addressing the third question and a review of the evidence on
beliefs.

\paragraph{The Null Hypothesis:}

Suppose that we observe $s$ shooters; each shoots $n$ consecutive
shots under identical conditions. Let $\mathbf{X}_{i}=\left\{ X_{ij}\right\} _{j=1}^{n}$
denote the vector of shot outcomes for shooter $i$ and $\mathbf{X}=\left\{ \mathbf{X}_{i}\right\} _{i=1}^{s}$
the matrix of these outcomes for all shooters, with $X_{ij}=1$ denoting
a made shot and $X_{ij}=0$ denoting a missed shot.

We would like to test the hypothesis that there is no positive serial
dependence in the outcomes of the observed shots. A test of the joint
null hypothesis 
\[
H_{0}:\text{ }\mathbf{X}_{i}\text{ is i.i.d. for each }i\text{ in }1,\ldots,s
\]
 assesses whether basketball shooting is a random process for all
shooters in the sample. In contrast, tests of the individual hypotheses

\[
H_{0}^{i}:\text{ }\mathbf{X}_{i}\text{ is i.i.d.,}
\]
or the multiple hypothesis problem that tests the hypotheses $H_{0}^{i}$
simultaneously assess whether basketball shooting is a random process
for shooter $i$ or for each of the shooters in the sample simultaneously,
respectively.\footnote{There is an ambiguity in the literature on the hot hand fallacy whether
belief in the hot hand refers to a biased belief in a serial correlation
or a causal effect of an outcome of a shot on subsequent shots. In
Online Appendix B.1, we characterize the relationship between the
null hypothesis that $\mathbf{X}_{i}$ is i.i.d.\ and the null hypothesis
that, for each shot, the outcomes of the preceding $m$ shots have
no causal effect on the probability of a make. We provide an unconfoundedness
type assumption under which these conditions are equivalent.}

Rejection of the joint null hypothesis $H_{0}$ indicates that there
is non-zero serial dependence for at least one shooter in the sample,
but does not indicate \textit{which} shooters deviate from randomness.
In order to identify any such shooters, we apply multiple testing
methods that control the familywise error rate (FWER), i.e., the probability
of at least one false rejection of an individual hypothesis $H_{0}^{i}$.
Note that a test of the joint null hypothesis $H_{0}$ is more liberal
than simultaneous tests of $H_{0}^{i}$, in the sense that tests of
$H_{0}$ can be rejected even if there is insufficient evidence to
reject any of the individual hypotheses $H_{0}^{i}$ at the same level.\footnote{Indeed, the closure method for constructing multiple tests that control
the FWER is based on tests of joint hypotheses; in order for $H_{0}^{i}$
to be rejected, tests of all joint null hypotheses for any subset
of shooters containing shooter $i$ must be rejected, not just the
subset consisting of all shooters. In fact, any multiple hypothesis
testing method that controls the FWER must be constructed with the
closure method \citep{romano2011consonance}.}

\paragraph{Streaky Alternatives:}

In general, stationary processes are a broad class of alternatives
to i.i.d.\ processes, allowing for quite arbitrary dependence. We
maintain the assumption that the shot outcomes $\mathbf{X}_{i}$ follow
stationary Bernoulli ($p_{i}$) processes $\mathbb{P}_{i}$ and are
independent across shooters, with $\mathbb{P}=\left\{ \mathbb{P}_{i}\right\} _{i=1}^{s}$
collecting these processes in a $2^{s}$-vector valued Bernoulli process.\footnote{Under $H_{0}$, $p_{i}=\mathbb{P}_{i}\left\{ X_{ij}=1\right\} $ may
vary across shooters.}

However, some stationary alternatives to $H_{0}$ are inconsistent
with notions of ``the hot hand'' or ``streak shooting''. GVT argue
that in most conceptions of the hot hand, the probability of making
a shot following a series of made shots is higher than both the marginal
probability of making a shot and the probability of making a shot
following a series of missed shots. Thus, following GVT and MS, in
order to assess whether there is positive serial dependence in basketball
shooting, we study tests of $H_{0}$ against alternatives in which
the parameters 
\begin{eqnarray}
\bar{\theta}{}_{P}^{k}\left(\mathbb{P}\right)=\frac{1}{s}\sum_{i=1}^{s}\theta_{P}^{k}\left(\mathbb{P}_{i}\right) & \text{ and } & \bar{\theta}{}_{D}^{k}\left(\mathbb{P}\right)=\frac{1}{s}\sum_{i=1}^{s}\theta_{D}^{k}\left(\mathbb{P}\right),\label{eq: theta bar}
\end{eqnarray}
where
\begin{flalign}
\theta_{P}^{k}\left(\mathbb{P}_{i}\right) & =\mathbb{P}_{i}\left\{ X_{i,j+k}=1\vert\prod_{l=0}^{k-1}X_{i,j+l}=1\right\} -\text{\ensuremath{\mathbb{P}}}_{i}\left\{ X_{ij}=1\right\} \label{eq: theta_P}\\
\theta_{D}^{k}\left(\text{\ensuremath{\mathbb{P}}}_{i}\right) & =\mathbb{P}_{i}\left\{ X_{i,j+k}=1\vert\prod_{l=0}^{k-1}X_{i,j+l}=1\right\} -\mathbb{P}_{i}\left\{ X_{i,j+k}=1\vert\prod_{l=0}^{k-1}\left(1-X_{i,j+l}\right)=1\right\} ,\label{eq: theta_D}
\end{flalign}
are greater than zero for some integer $k$. Throughout, we refer
to alternatives of this form as ``streaky'', as the probability
of a streak of made shots of length $k+1$ is higher than it would
be under an i.i.d.\ process. A rejection of the null hypothesis $H_{0}$
against streaky alternatives provides an affirmative answer to the
first question posed at the beginning of this section -- that there
is non-zero and positive serial dependence in basketball shooting.

\paragraph{Test Statistics:}

Following GVT and MS, we study tests of the individual and joint null
hypotheses of randomness against streaky alternatives that use natural
plug-in estimators for the parameters $\theta_{P}^{k}\left(\mathbb{P}_{i}\right)$
and $\theta_{D}^{k}\left(\text{\ensuremath{\mathbb{P}}}_{i}\right)$,
as well as $\bar{\theta}{}_{P}^{k}\left(\mathbb{P}\right)$ and $\bar{\theta}_{D}^{k}\left(\text{\ensuremath{\mathbb{P}}}_{i}\right)$,
as test statistics, respectively. These statistics are defined as
follows. Let each individual's observed shooting percentage be given
by $\hat{p}_{n,i}=\frac{1}{n}\sum_{j=1}^{n}X_{ij}$ and let $\hat{P}_{n,k}(\mathbf{X}_{i})$
denote the proportion of made shots following $k$ consecutive made
shots. That is, letting $Y_{ijk}=\prod_{l=j}^{j+k}X_{il}$ and $V_{ik}=\sum_{j=1}^{n-k}Y_{ijk}$,
then $\hat{P}_{n,k}(\mathbf{X}_{i})$ is given by

\begin{equation}
\hat{P}_{n,k}(\mathbf{X}_{i})=\frac{V_{ik}}{V_{i\left(k-1\right)}}.\label{eq:P_k}
\end{equation}
Likewise, let $\hat{D}_{n,k}\left(\mathbf{X}_{i}\right)$ denote the
difference between the proportion of made shots following $k$ consecutive
made shots and $k$ consecutive missed shots. That is, letting $Z_{ijk}=\prod_{l=j}^{j+k}\left(1-X_{il}\right)$
and $W_{ik}=\sum_{j=1}^{n-k}Z_{ijk}$, then $\hat{D}_{n,k}(\mathbf{X}_{i})$
is given by
\begin{equation}
\hat{D}_{n,k}\left(\mathbf{X}_{i}\right)=\frac{V_{ik}}{V_{i\left(k-1\right)}}-\frac{W_{ik}}{W_{i\left(k-1\right)}}\label{eq:D_k}
\end{equation}
Intuitively, $\hat{P}_{n,k}(\mathbf{X}_{i})-\hat{p}_{n,i}$ and $\hat{D}_{n,k}\left(\mathbf{X}_{i}\right)$
are natural plug-in estimators for $\theta_{P}^{k}\left(\mathbb{P}_{i}\right)$
and $\theta_{D}^{k}\left(\text{\ensuremath{\mathbb{P}}}_{i}\right)$,
respectively. The averages of these estimators over the shooters in
the sample are denoted by
\begin{eqnarray}
\bar{P}_{k}\left(\mathbf{X}\right)=\frac{1}{s}\sum_{i=1}^{s}\hat{P}_{n,k}\left(\mathbf{X}_{i}\right)-\hat{p}_{n,i} & \text{ and } & \bar{D}_{k}\left(\mathbf{X}\right)=\frac{1}{s}\sum_{i=1}^{s}\hat{D}_{n,k}\left(\mathbf{X}_{i}\right)\label{eq: averages}
\end{eqnarray}
and are natural plug-in estimators for $\bar{\theta}{}_{P}^{k}\left(\mathbb{P}\right)$
and $\bar{\theta}{}_{D}^{k}\left(\mathbb{P}\right)$, respectively.
Note that $\hat{P}_{n,k}(\mathbf{X}_{i})$ and $\hat{D}_{n,k}(\mathbf{X}_{i})$
are not defined for every sequence $\mathbf{X}_{i}$. Specifically,
they are not defined for sequences without instances of $k$ consecutive
ones or zeros. However, under our null hypothesis of randomness and
the alternatives that we consider, the statistics are defined with
probability approaching one exponentially quickly as $n$ grows to
infinity.

\paragraph{Estimation:}

Rejection of either the joint null hypothesis $H_{0}$ or an individual
hypothesis $H_{0}^{i}$, after accounting for simultaneity, indicates
that there is non-zero serial dependence for at least one shooter
in the sample. It does not, however, provide a quantification of this
dependence. Estimates and confidence intervals for $\bar{\theta}{}_{P}^{k}\left(\mathbb{P}\right)$
and $\bar{\theta}{}_{D}^{k}\left(\mathbb{P}\right)$ provide metrics
for quantifying the observed serial dependence. These metrics are
adopted by GVT and MS. As suggested above, $\bar{P}_{k}\left(\mathbf{X}\right)$
and $\bar{D}_{k}\left(\mathbf{X}\right)$ are natural estimators for
$\bar{\theta}{}_{P}^{k}\left(\mathbb{P}\right)$ and $\bar{\theta}{}_{D}^{k}\left(\mathbb{P}\right)$,
respectively. While not the emphasis of our analysis, we discuss methods
for constructing confidence intervals for these parameters in Section
\ref{subsec:Asymptotic-Test}.
\begin{singlespace}

\section{Testing Randomness Against Streaky Alternatives\label{sec:Detecting}}
\end{singlespace}

\noindent In this Section, we develop methods for testing the randomness
of a collection of Bernoulli sequences against streaky alternatives
using the plug-in statistics presented in Section \ref{sec:Posing-the-Problem}.

\medskip{}

\subsection{Asymptotic Behavior of the Test Statistics\label{subsec:Asymptotic-Test}}

We begin by characterizing the asymptotic distributions of the plug-in
statistics $\hat{P}_{n,k}(\mathbf{X}_{i})-\hat{p}_{n}$ and $\hat{D}_{n,k}(\mathbf{X}_{i})$
under the null hypothesis of randomness $H_{0}^{i}$. To date, such
distributions have not been derived. \citet{miller2018cold} claim
that $\hat{P}_{n,k}(\mathbf{X}_{i})$ is asymptotically normal under
the null hypothesis, referencing \citet{mood1940distribution}, but
do not provide explicit formulae for its asymptotic variance. Even
in the i.i.d.\ case, the test statistics are functions of overlapping
subsequences of observations, so central limit theorems for dependent
data are required.
\begin{thm}
\label{thm: Convergence_under_H0}Under the assumption that $\mathbf{X}_{i}=\left\{ X_{ij}\right\} _{j=1}^{n}$
is a sequence of i.i.d.\ Bernoulli($p_{i}$) random variables,

\noindent \textbf{(i)} $\hat{P}_{n,k}\left(\mathbf{X}_{i}\right)-\hat{p}_{n,i,}$,
with $\hat{P}_{n,k}\left(\mathbf{X}_{i}\right)$ given by (\ref{eq:P_k})
and $\hat{p}_{n,i}=n^{-1}\sum_{j=1}^{n}X_{ij}$, is asymptotically
normal with limiting distribution given by
\begin{equation}
\sqrt{n}\left(\hat{P}_{n,k}\left(\mathbf{X}_{i}\right)-\hat{p}_{n,i}\right)\overset{d}{\rightarrow}N\left(0,\sigma_{P}^{2}\left(p_{i},k\right)\right),\label{eq: P_k conv}
\end{equation}
as $n\to\infty$, where \textup{$\sigma_{P}^{2}\left(p_{i},k\right)=p_{i}^{1-k}\left(1-p_{i}\right)\left(1-p_{i}^{k}\right)$},
and

\noindent \textbf{(ii)} $\hat{D}_{n,k}\left(\mathbf{X}_{i}\right)$,
given by (\ref{eq:D_k}), is asymptotically normal with limiting distribution
given by

\noindent 
\begin{equation}
\sqrt{n}\hat{D}_{n,k}\left(\mathbf{X}_{i}\right)\overset{d}{\rightarrow}N\left(0,\sigma_{D}^{2}\left(p_{i},k\right)\right),\label{eq: D_k conv}
\end{equation}
as $n\to\infty$, where \textup{$\sigma_{D}^{2}\left(p_{i},k\right)=\left(p_{i}\left(1-p_{i}\right)\right)^{1-k}\left(\left(1-p_{i}\right)^{k}+p_{i}^{k}\right)$.}
\end{thm}
\begin{rem}
Note that $\sigma_{D}^{2}\left(\frac{1}{2},k\right)=2^{k-1}$ increases
exponentially with $k$, stemming from an effectively reduced sample
size -- limiting to those outcomes that follow sequences of ones
or zeros of length $k$. $\hfill\blacksquare$
\end{rem}
\begin{rem}
Theorem \ref{thm: Convergence_under_H0} can be generalized to a triangular
array $\mathbf{X}_{n,i}=\left\{ X_{n,i,j}\right\} _{j=1}^{n}$ of
i.i.d.\ Bernoulli trials with probability of success $p_{n,i}$ converging
to $p_{i}$. Specifically, we have that, under $p_{n,i}$ (\ref{eq: P_k conv})
and (\ref{eq: D_k conv}) continue to hold. This result implies that
we can consistently approximate the quantiles of the distributions
of $\hat{P}_{n,k}\left(\mathbf{X}_{n,i}\right)-\hat{p}_{n,i}$ and
$\hat{D}_{n,k}\left(\mathbf{X}_{n,i}\right)$ under the null hypothesis
with the parametric bootstrap, which approximates the distribution
of $\sqrt{n}\hat{D}_{n,k}\left(\mathbf{X}_{i}\right)$ under $p_{i}$
using that of $\sqrt{n}\hat{D}_{n,k}\left(\mathbf{X}_{i}\right)$
under $\hat{p}_{n,i}$.$\hfill\blacksquare$
\end{rem}
\begin{rem}
For a set $\mathbf{X}=\left\{ \mathbf{X}_{i}\right\} _{i=1}^{s}$
of Bernoulli sequences of length $n$ each having probability of success
$p_{i}$, Theorem \ref{thm: Convergence_under_H0} implies that the
statistics $\sqrt{ns}\bar{P}_{k}\left(\mathbf{X}\right)$ and $\sqrt{ns}\bar{D}_{k}\left(\mathbf{X}\right)$
have normal limiting distributions with means equal to zero and variances
equal to the averages of $\sigma_{P}^{2}\left(p_{i},k\right)$ and
$\sigma_{D}^{2}\left(p_{i},k\right)$ over the individuals $1,\ldots,s$,
respectively.\footnote{In Online Appendix C, we discuss the asymptotic distributions of these
statistics when the parameters $p_{i}$ are realizations of a sequence
of i.i.d.\ random variables. We use this result in Section \ref{sec:Power}
to approximate the power of the tests that we develop against a set
of alternatives in which individuals independently deviate from the
null hypothesis with a pre-specified probability.} $\hfill\blacksquare$
\end{rem}
\begin{rem}
In Online Appendix F, we show that under a stationary $\alpha$-mixing
process $\mathbb{P}_{i}$, $\hat{P}_{n,k}\left(\mathbf{X}_{i}\right)-\hat{p}_{n,i}$
and $\hat{D}_{n,k}\left(\mathbf{X}_{i}\right)$ are asymptotically
normal with means $\theta_{P}^{k}\left(\text{\ensuremath{\mathbb{P}}}_{i}\right)$
and $\theta_{D}^{k}\left(\text{\ensuremath{\mathbb{P}}}_{i}\right)$,
respectively. Stationary $\alpha$-mixing processes provide a general
class of alternatives to the null hypothesis of randomness, and allow
for quite general forms of dependence between shots that are close
to each other.\footnote{In an $\alpha$-mixing process, the dependence between two shots $X_{ij}$
and $X_{i\left(j+t\right)}$ approaches zero as $t$ grows to infinity.
For example, any Markov Chain with finite state space that is irreducible
and aperiodic is $\alpha$-mixing \citep{bradley1986basic}. In Online
Appendix B.2, we show that stationary$\alpha$-mixing alternatives
are a natural class of alternatives to consider in a dynamic potential
outcomes framework.} $\hfill\blacksquare$
\end{rem}
\begin{rem}
For a stationary process $\mathbb{P}_{i}$, the limiting variances
of $\hat{P}_{n,k}\left(\mathbf{X}_{i}\right)-\hat{p}_{n,i}$ and $\hat{D}_{n,k}\left(\mathbf{X}_{i}\right)$
can be quite complicated. However, they, as well as their entire sampling
distributions, can be estimated with general bootstrap methods for
stationary time series (see \citet{lahiri2013resampling}), such as
the moving blocks bootstrap \citep{liu1992moving,kunsch1989jackknife},
the stationary bootstrap \citep{politis1994stationary}, or subsampling
\citep{politis1999subsampling}. Such methods provide asymptotically
valid confidence intervals for general parameters, such as $\theta_{P}^{k}\left(\text{\ensuremath{\mathbb{P}}}_{i}\right)$
or $\theta_{D}^{k}\left(\text{\ensuremath{\mathbb{P}}}_{i}\right)$.
$\hfill\blacksquare$
\end{rem}
There is a severe second-order bias in the finite sample performance
of these asymptotic approximations. Specifically, let $\beta_{P}^{n,k}\left(\mathbb{P}_{i}\right)$
and $\beta_{D}^{n,k}\left(\mathbb{P}_{i}\right)$ denote the expectations
of $\hat{P}_{n,k}\left(\mathbf{X}_{i}\right)-\hat{p}_{n,i}$ and $\hat{D}_{n,k}\left(\mathbf{X}_{i}\right)$
under the stationary process $\mathbb{P}_{i}$. With a minor abuse
of notation, we let $\beta_{P}^{n,k}\left(p_{i}\right)$ and $\beta_{D}^{n,k}\left(p_{i}\right)$
denote these parameters when $\mathbb{P}_{i}$ is an i.i.d.\ Bernoulli
process with marginal success rate $p_{i}$. MS show that $\beta_{P}^{n,k}\left(p_{i}\right)$
and $\beta_{D}^{n,k}\left(p_{i}\right)$ are less than $\theta_{P}^{k}\left(\text{\ensuremath{\mathbb{P}}}_{i}\right)$
and $\theta_{D}^{k}\left(\text{\ensuremath{\mathbb{P}}}_{i}\right)$
under $H_{0}^{i}$ with marginal success rate $p_{i}$. These differences
converge to zero as $n$ increases.\footnote{Exact expressions for the expectations of these statistics in finite-samples
appear to be unknown for $k>1$. In Online Appendix D, we obtain the
second order approximations $\beta_{P}^{n,k}\left(p_{i}\right)=n^{-1}p_{i}\left(1-p_{i}^{-k}\right)+O\left(n^{-2}\right)$
and $\beta_{D}^{n,k}\left(p_{i}\right)=n^{-1}\left(1-\left(1-p_{i}\right)^{1-k}-p_{i}^{1-k}\right)+O\left(n^{-2}\right).$}

Thus, the statistics $\hat{P}_{n,k}\left(\mathbf{X}_{i}\right)-\hat{p}_{n,i}$
or $\hat{D}_{n,k}\left(\mathbf{X}_{i}\right)$ have a negative bias
when considered as estimators for $\theta_{P}^{k}\left(\mathbb{P}_{i}\right)$
and $\theta_{D}^{k}\left(\text{\ensuremath{\mathbb{P}}}_{i}\right)$
under $H_{0}^{i}$. Equivalently, procedures that test $H_{0}^{i}$
against streaky alternatives by comparing $\hat{P}_{n,k}\left(\mathbf{X}_{i}\right)-\hat{p}_{n,i}$
or $\hat{D}_{n,k}\left(\mathbf{X}_{i}\right)$ to quantiles of their
limiting distributions -- without correcting for these finite-sample
biases -- have a type 1 error rate below the desired level in finite-samples.
To illustrate, suppose that $n=100$ and that $\mathbf{X}_{i}=\left\{ X_{ij}\right\} _{j=1}^{n}$
is an i.i.d.\ Bernoulli$(p_{i})$ sequence with $p_{i}=1/2$. Columns
(1) and (3) in Table \ref{tab:bias-level} give the expectations of
$\hat{P}_{n,k}(\mathbf{X}_{i})-\hat{p}_{n,i}$ and $\hat{D}_{n,k}(\mathbf{X}_{i})$
for $k$ in $1,\dotsc,4$, respectively. In contrast, $\theta_{P}^{k}\left(\text{\ensuremath{\mathbb{P}}}_{i}\right)$
and $\theta_{D}^{k}\left(\text{\ensuremath{\mathbb{P}}}_{i}\right)$,
defined in (\ref{eq: theta_P}) and (\ref{eq: theta_D}), are both
equal to zero. Columns (2) and (4) of Table \ref{tab:bias-level}
give the probabilities that $\hat{P}_{n,k}(\mathbf{X}_{i})-\hat{p}_{n,i}$
and $\hat{D}_{n,k}(\mathbf{X}_{i})$ are greater than the $0.95$
quantiles of the normal distributions with means zero and variances
$\sigma_{D}^{2}\left(p_{i},k\right)$ and $\sigma_{P}^{2}\left(p_{i},k\right)$
for $k$ in $1,\dotsc,4$, respectively. The probabilities are significantly
below $0.05$ and decrease with $k$. Hence, to conduct more powerful
tests of randomness, it is necessary to account for this bias --
at least implicitly.

\begin{table}[t]
\begin{centering}
\begin{tabular}{ccccccc}
\toprule 
 &  & \multicolumn{2}{c}{$\hat{P}_{n,k}(\mathbf{X}_{i})-\hat{p}_{n,i}$} &  & \multicolumn{2}{c}{$\hat{D}_{n,k}(\mathbf{X}_{i})$}\tabularnewline
\cmidrule{3-4} \cmidrule{4-4} \cmidrule{6-7} \cmidrule{7-7} 
$k$ &  & Expectation & Type 1 Error Rate &  & Expectation & Type 1 Error Rate\tabularnewline
\cmidrule{1-1} \cmidrule{3-4} \cmidrule{4-4} \cmidrule{6-7} \cmidrule{7-7} 
 &  & (1) & (2) &  & (3) & (4)\tabularnewline
\midrule
\midrule 
1 &  & -0.005 & 0.044 &  & -0.010 & 0.039\tabularnewline
\midrule
2 &  & -0.016 & 0.032 &  & -0.032 & 0.029\tabularnewline
\midrule
3 &  & -0.041 & 0.023 &  & -0.080 & 0.020\tabularnewline
\midrule
4 &  & -0.090 & 0.013 &  & -0.177 & 0.010\tabularnewline
\bottomrule
\end{tabular}
\par\end{centering}
\medskip{}

\begin{centering}
\caption{\label{tab:bias-level}Finite-Sample Behavior of Plug-in Statistics}
\par\end{centering}
\medskip{}

{\footnotesize{}Notes: Table displays simulated estimates of the finite
sample expectations of $\hat{P}_{n,k}(\mathbf{X}_{i})-\hat{p}_{n,i}$
and $\hat{D}_{n,k}(\mathbf{X}_{i})$ as well as the type 1 error rates
of the hypothesis tests that reject $H_{0}^{i}$ if $\hat{P}_{n,k}(\mathbf{X}_{i})-\hat{p}_{n,i}$
and $\hat{D}_{n,k}(\mathbf{X}_{i})$ exceed the $0.95$ quantile of
their asymptotic distributions. We take 100,000 draws of Bernoulli($1/2)$
random variables of length $100$. We compute expectations by taking
the mean of $\hat{P}_{n,k}(\mathbf{X}_{i})-\hat{p}_{n,i}$ and $\hat{D}_{n,k}(\mathbf{X}_{i})$
computed on each draw. We compute type 1 error rates by taking the
proportion of draws in which $\hat{P}_{n,k}(\mathbf{X}_{i})-\hat{p}_{n,i}$
and $\hat{D}_{n,k}(\mathbf{X}_{i})$ exceed the $0.95$ quantiles
of their asymptotic distributions.}{\footnotesize\par}
\end{table}

In their analysis of controlled basketball shooting experiments, GVT
test the individual hypotheses $H_{0}^{i}$ by comparing $\hat{P}_{n,k}\left(\mathbf{X}_{i}\right)-\hat{p}_{n,i}$
and $\hat{D}_{n,k}\left(\mathbf{X}_{i}\right)$ to quantiles of approximations
to their limiting distributions without correcting for finite-sample
bias. MS argue that GVT's conclusion that the null hypotheses $H_{0}^{i}$
cannot be rejected is sensitive to correction for this bias, i.e.,
the implementation of tests with more accurate control of the type
1 error rate. In the subsequent subsection, we discuss permutation
tests, show that they automatically account for finite-sample biases,
and prove that they are in fact the only tests that control the type
1 error rate exactly in finite samples. We advocate for their choice
as the default test in our setting.

\subsection{Permutation Tests, Bias-Corrected Estimation, and Simultaneous Inference\label{subsec:Permutation-Tests}}

In this subsection, we outline permutation tests of the individual
hypothesis $H_{0}^{i}$ and the joint null hypothesis $H_{0}$ that
control the type 1 error rate exactly in finite-samples. We then propose
a set of estimators of the individual parameters $\theta_{P}^{k}\left(\mathbb{P}_{i}\right)$
and $\theta_{D}^{k}\left(\text{\ensuremath{\mathbb{P}}}_{i}\right)$
and the joint parameters $\bar{\theta}{}_{P}^{k}\left(\mathbb{P}\right)$
and $\theta_{D}^{k}\left(\text{\ensuremath{\mathbb{P}}}_{i}\right)$
that are exactly unbiased under the null hypothesis. Finally, we lay
out a standard multiple hypothesis testing procedure that can be applied
to test the individual hypotheses $H_{0}^{i}$ simultaneously.

\paragraph{Individual and Joint Tests:}

Based on the data $\mathbf{X}_{i}=\left\{ X_{ij}\right\} _{j=1}^{n}$,
it is desired to test the null hypothesis $H_{0}^{i}$ that the underlying
observations are i.i.d.\ Bernoulli with some unknown success probability
$p_{i}$. Under $H_{0}^{i}$, the distribution of $\mathbf{X}_{i}$
is invariant under permutations; that is $\left(X_{i,1},\ldots,X_{i,n}\right)$
and $\left(X_{i,\pi\left(1\right)},\ldots,X_{i,\pi\left(n\right)}\right)$,
where $\pi$ is a permutation of $\left(1,\ldots,n\right)$, have
the same joint distribution. This property is a special case of the
\textit{randomization hypothesis }specified in Section 15.2 of \citet{lehmann2005testing}
and allows for the construction of permutation tests.

In a permutation test, a test statistic is recomputed on every permutation
of a data set. The distribution of these recomputed statistics is
used as a null or reference distribution for comparison with the observed
value of the test statistic. The proportion of recomputed statistics
exceeding the observed test statistic is the $p$-value of the permutation
test. Permutation tests are exact level $\alpha$ for any choice of
test statistic. In particular, let $T_{n}\left(\mathbf{X}_{i}\right)$
be any real-valued test statistic for testing $H_{0}^{i}$. Let $\mathbf{X}_{i,\pi}=\left(X_{i,\pi\left(1\right),\ldots,}X_{i,\pi\left(n\right)}\right),$
where $\pi$ is an element of $\Pi\left(n\right)$, be the set of
permutations of $\left\{ 1,\ldots,n\right\} $. The permutation, or
randomization, distribution for $\sqrt{n}T_{n}\left(\mathbf{X}_{i}\right)$
is given by 
\[
\hat{R}_{n}^{T}\left(t\right)=\frac{1}{n!}\sum_{\pi\in\Pi\left(n\right)}I\left\{ \sqrt{n}T_{n}\left(\mathbf{X}_{i,\pi}\right)\leq t\right\} .
\]
For a nominal level $\alpha$, $0<\alpha<1$, the permutation test
rejects at level $\alpha$ if $\sqrt{n}T_{n}\left(\mathbf{X}_{i}\right)$
is greater than the $1-\alpha$ quantile of $\hat{R}_{n}^{T}$.\footnote{As the permutation distribution is discrete, the exact permutation
test may require randomization when $\sqrt{n}T_{n}\left(\mathbf{X}_{i}\right)$
is equal to the $1-\alpha$ quantile of $\hat{R}_{n}^{T}$. In practice,
we use a slightly conservative approach by not randomizing; that is,
we reject $H_{0}^{i}$ only if $\sqrt{n}T_{n}\left(\mathbf{X}_{i}\right)$
exceeds the $1-\alpha$ quantile of $\hat{R}_{n}^{T}$.} Define the permutation test function $\varphi\left(\mathbf{X}_{i}\right)$
to be equal to one if the permutation test rejects and zero otherwise.
By Theorem 15.2.1 in \citet{lehmann2005testing}, $\mathbb{E}\left[\varphi\left(\mathbf{X}_{i}\right)\right]=\alpha$
if $H_{0}^{i}$ is true. What may be less obvious is that any test
$\varphi$ that is exactly level $\alpha$ for testing $H_{0}^{i}$
\textit{must} be a permutation test.
\begin{thm}
\label{thm: exact permutation}Suppose $\varphi=\varphi\left(\mathbf{X}_{i}\right)$
is any test function such that $\mathbb{E}\left[\varphi\left(\mathbf{X}_{i}\right)\right]=\alpha$
whenever $\mathbf{X}_{i}$ is i.i.d.\ Bernoulli with some unknown
success rate $p_{i}$. Then, $\varphi$ must be a permutation test;
that is 
\[
\frac{1}{n!}\sum_{\pi\in\Pi\left(n\right)}\varphi\left(\mathbf{X}_{i,\pi}\right)=\alpha.
\]
\end{thm}
In practice, one does not need to compute all $n!$ permutations.
Instead, if permutations are sampled at random, then one can still
attain valid finite-sample $p$-values. Both $\hat{P}_{n,k}\left(\mathbf{X}_{i}\right)-\hat{p}_{n,i}$
and $\hat{D}_{n,k}\left(\mathbf{X}_{i}\right)$, given in (\ref{eq:P_k})
and (\ref{eq:D_k}), are appropriate choices for $T_{n}\left(\mathbf{X}_{i}\right)$.
In the following subsection, we characterize the asymptotic distribution
of $\hat{R}_{n}^{T}$ for these choices.

Similarly, the joint null hypothesis $H_{0}$ can be tested with a
stratified permutation test wherein each Bernoulli sequence $\mathbf{X}_{i}$
is permuted separately. Specifically, let $K_{n,s}\left(\mathbf{T}\right)$
be a general function of the individual test statistics $\mathbf{T}=\left\{ T_{n}\left(\mathbf{X}_{i}\right)\right\} _{i=1}^{s}$.
The stratified permutation distribution for $\sqrt{ns}K_{n,s}$ is
given by

\[
\hat{R}_{n,s}^{K,T}\left(t\right)=\frac{1}{\left(n!\right)^{s}}\sum_{\left(\pi_{1},\ldots,\pi_{s}\right)\in\Pi\left(n\right)^{s}}I\left\{ \sqrt{ns}K_{n,s}\left(T_{n}\left(\mathbf{X}_{i,\pi_{i}}\right),\ldots,T_{n}\left(\mathbf{X}_{i,\pi_{s}}\right)\right)\leq t\right\} ,
\]
where $\Pi\left(n\right)^{s}$ is the set of all $s$-vectors of permutations
of $\left(1,\ldots,n\right)$.\footnote{We note that $K_{n,s}$ must be computed over all individuals $i$
where the statistic $T_{n}\left(\mathbf{X}_{i}\right)$ is defined.} A stratified permutation test rejects $H_{0}$ at level $\alpha$
if $\sqrt{ns}K_{n,s}$ exceeds the $1-\alpha$ quantile of $\hat{R}_{n,s}^{K,T}$.
Both $\bar{P}_{k}\left(\mathbf{X}\right)$ and $\bar{D}_{k}\left(\mathbf{X}\right)$,
given in (\ref{eq: averages}), are appropriate choices for the joint
test statistic $K_{n,s}$.\footnote{In Online Appendix G, we outline three additional choices for joint
test statistics that combine $p$-values of individual permutation
tests across individuals. Each of these choices of statistics will
have power against different alternatives. Additionally, we outline
two methods for combining $p$-values of different joint tests to
compute a singular composite $p$-value.}

The use of permutation tests bypasses the need for explicit bias-correction.
Specifically, the expected value of the mean of the permutation distribution
$\hat{R}_{n}^{T}\left(t\right)$ is exactly that of $\sqrt{n}T_{n}\left(\mathbf{X}_{i}\right)$
under the null hypothesis. Thus, one can avoid approximating finite
sample biases explicitly, because the permutation distributions account
for these biases automatically.

\paragraph{Bias-Corrected Estimation:}

The equality between expectations of the means of permutation distributions
and expectations of their associated statistics under the null hypothesis
can be leveraged to construct bias-corrected estimators. In particular,
let 
\[
\hat{\eta}\left(\mathbf{X}_{i},T_{n}\right)=\frac{1}{n!}\sum_{\pi\in\Pi\left(n\right)}T_{n}\left(\mathbf{X}_{i,\pi}\right)
\]
denote the mean of the permutation distribution of the statistic $T_{n}\left(\mathbf{X}_{i}\right)$.
Under the null hypothesis, the expectation of $\hat{\eta}\left(\mathbf{X}_{i},T_{n}\right)$
is exactly equal to the expectation of $T_{n}\left(\mathbf{X}_{i}\right)$.
This observation suggests the bias-corrected estimators
\begin{equation}
\tilde{P}_{n,k}(\mathbf{X}_{i})=\hat{P}_{n,k}(\mathbf{X}_{i})-\hat{p}_{n,i}-\hat{\eta}\left(\mathbf{X}_{i},\hat{P}_{k}-\hat{p}_{i}\right)\text{ and }\tilde{D}_{n,k}(\mathbf{X}_{i})=\hat{D}_{n,k}(\mathbf{X}_{i})-\hat{\eta}\left(\mathbf{X}_{i},\hat{D}_{n,k}\right)\label{eq: corrected}
\end{equation}
and their averages 
\begin{equation}
\bar{\tilde{P}}_{k}\left(\mathbf{X}\right)=\frac{1}{s}\sum_{i=1}^{s}\tilde{P}_{n,k}(\mathbf{X}_{i})\text{ and }\bar{\tilde{D}}_{k}\left(\mathbf{X}\right)=\frac{1}{s}\sum_{i=1}^{s}\tilde{D}_{n,k}\left(\mathbf{X}_{i}\right).\label{eq: corrected averages}
\end{equation}
These estimators are exactly unbiased under the null hypothesis, and
are consistent under streaky alternatives.\footnote{Alternatively, the bias can be approximated by the parametric bootstrap,
i.e., $\hat{P}_{n,k}(\mathbf{X}_{i})-\hat{p}_{n,i}-\beta_{P}^{n,k}\left(\hat{p}_{n,i}\right)$
and $\hat{D}_{n,k}(\mathbf{X}_{i})-\beta_{D}^{n,k}\left(\hat{p}_{n,i}\right)$
where $\beta_{P}^{n,k}\left(\hat{p}_{n,i}\right)$ and $\beta_{D}^{n,k}\left(\hat{p}_{n,i}\right)$
are computed with simulation. MS take this approach. These estimators
are only approximately unbiased under the null hypothesis. It is straightforward
to show that the expectations of these statistics are $O\left(n^{-2}\right)$
by replacing $p_{i}$ with $\hat{p}_{n,i}$ in the second order approximations
given in Online Appendix D.}

\paragraph{Simultaneous Tests:}

Suppose that the joint null hypothesis $H_{0}$ is rejected. In this
case, in order to characterize which of the Bernoulli sequences $\mathbf{X}_{i}$
are non-random, we would like to know which of the individual hypothesis
$H_{0}^{i}$ can be rejected. A problem of this form -- testing a
finite number of individual hypotheses simultaneously -- is a ``multiple
testing'' or ``simultaneous inference'' problem; see Chapter 9
of \citet{lehmann2005testing} for a textbook treatment.

If the hypotheses $H_{0}^{i}$ are each tested at level $\alpha$,
then the probability of a false rejection of at least one individual
hypothesis $H_{0}^{i}$ increases rapidly with $s$. In fact, when
$s$ is equal to $10$, then the probability of at least one false
rejection when all individual hypotheses are true is equal to approximately
$0.4$. Thus, we apply methods that control the familywise error rate
(FWER), i.e., the probability of at least one false rejection of an
individual null hypothesis $H_{0}^{i}$.

In particular, we apply a stepdown procedure with \v{S}id\'{a}k
critical values. Let $\rho_{i}$ denote the $p$-value for a permutation
test of $H_{0}^{i}$ and let the $p$-values ordered from lowest to
highest be $\rho_{(1)},\ldots,\rho_{(s)}$, with associated hypotheses
$H_{0}^{(1)},\ldots,H_{0}^{(s)}$. Fix a nominal level $\alpha$,
$0<\alpha<1$, and let $r$ be the maximal index such that $\rho_{(1)}<\alpha_{1},\cdots,\rho_{(r)}<\alpha_{r}$
and $\rho_{(r+1)}>\alpha_{r+1}$, where
\[
\alpha_{i}=1-\left(1-\alpha\right)^{\left(1/\left(s-i+1\right)\right)},
\]
Then, if the tests of $H_{0}^{i}$ are independent, the stepdown procedure
with \v{S}id\'{a}k critical values rejects the hypotheses $H_{0}^{(1)},\ldots,H_{0}^{(r)}$
and has FWER less than or equal to $\alpha$. If the tests of $H_{0}^{i}$
are independent, as they are in our application to controlled basketball
shooting experiments, then the stepdown procedure with \v{S}id\'{a}k
critical values is optimal in a maximin sense; see Section 9.2 of
\citet{lehmann2005testing} for a detailed discussion.\footnote{For cases where tests of $H_{0}^{i}$ are not independent, the stepdown
method of \citet{romano2005exact} can be applied. The first step
of this procedure can be used as a test of the joint null hypothesis
$H_{0}$.}

\subsection{Asymptotic Behavior of the Permutation Distributions\label{subsec:Asymptotic-Permutation}}

In this section, we describe the limiting behavior of the permutation
distributions of\\
 $\sqrt{n}\left(\hat{P}_{n,k}\left(\mathbf{X}_{i}\right)-\hat{p}_{n,i}\right)$
and $\sqrt{n}\hat{D}_{n,k}\left(\mathbf{X}_{i}\right)$ under the
null hypothesis that $\mathbf{X}_{i}$ is i.i.d.\ and under general
stationary alternatives. In Section \ref{sec:Power}, these results
allow us to study the power of the permutation tests outlined in Section
\ref{subsec:Permutation-Tests} against particular stationary alternatives.
We are aided by an appropriate central limit theorem using Stein's
method (see Rinott 1994 and Stein 1986). The permutation distribution
itself is random, but depends only on the number of ones in $\mathbf{X}_{i}$,
which, under i.i.d.\ sampling, is binomial.
\begin{thm}
\label{thm: Null Permutation}Under the assumption that $\mathbf{X}_{i}=\left\{ X_{ij}\right\} _{j=1}^{\infty}$
are $i.i.d$ Bernoulli ($p_{i}$) variables, then

\noindent \textbf{(i)} the permutation distribution of $\sqrt{n}T_{n}$
based on the test statistic $T_{n}=\hat{D}_{n,k}\left(X_{i1},\dots,X_{in}\right)$
satisfies
\[
\sup_{t}\vert\hat{R}_{n}^{T}\left(t\right)-\Phi\left(t/\sigma_{D}\left(p_{i},k\right)\right)\vert\overset{P}{\rightarrow}0
\]

\noindent as $n\to\infty$, where $\overset{P}{\to}$ denotes convergence
in probability and $\Phi\left(\cdot\right)$ denotes the standard
normal cumulative distribution function, and

\noindent \textbf{(ii)} the permutation distribution of $\sqrt{n}T_{n}$
based on the test statistic $T_{n}=\hat{P}_{n,k}\left(X_{i1},\dots,X_{in}\right)-\hat{p}_{n,i}$
satisfies
\[
\sup_{t}\vert\hat{R}_{n}^{T}\left(t\right)-\Phi\left(t/\sigma_{P}\left(p_{i},k\right)\right)\vert\overset{P}{\rightarrow}0
\]
as $n\to\infty$, where $\sigma_{P}\left(p_{i},k\right)$ and $\sigma_{D}\left(p_{i},k\right)$
are given in Theorem \ref{thm: Convergence_under_H0}.
\end{thm}
Next, we study the limiting behavior of the permutation distributions
of $\sqrt{n}\left(\hat{P}_{n,k}\left(\mathbf{X}_{i}\right)-\hat{p}_{n,i}\right)$
and $\sqrt{n}\hat{D}_{n,k}\left(\mathbf{X}_{i}\right)$ in possibly
non-i.i.d.\ settings. We provide the details of this argument for
$\sqrt{n}\hat{D}_{n,1}\left(\mathbf{X}_{i}\right)$ and note that
the argument generalizes to $\sqrt{n}\hat{D}_{n,k}\left(\mathbf{X}_{i}\right)$
or $\sqrt{n}\left(\hat{P}_{n,k}\left(\mathbf{X}_{i}\right)-\hat{p}_{n,i}\right)$
for general $k$. We begin by considering the behavior of the permutation
distribution of $\sqrt{n}\hat{D}_{n,1}\left(\mathbf{X}_{i}\right)$
for fixed (nonrandom) sequences of the number of ones in $n$ Bernoulli
trials. In this case, the permutation distribution is not random,
but deriving its limiting behavior is nontrivial and requires the
application of a novel equicontinuity argument. In particular, let
$L_{n}\left(h\right)$ be the permutation distribution for $\sqrt{n}T_{n}$
based on a data set of length $n$ with 
\[
a_{n}=a_{n}\left(h\right)=\lfloor\frac{n}{2}+h\sqrt{n}\rfloor
\]
ones and $n-a_{n}$ zeros, where $\lfloor x\rfloor$ denotes the largest
integer less than or equal to $x$ . Observe that if $a_{n}$ is an
integer between $0$ and $n$, then $h=n^{-1/2}\left(a_{n}-\frac{n}{2}\right)$.
In Online Appendix Lemma K.2, we show that under nonrandom sequences
$h_{n}\to h$ and for $T_{n}=\hat{D}_{n,1}\left(\mathbf{X}_{i}\right)$,
we have that $L_{n}\left(h_{n}\right)\overset{d}{\to}N\left(0,1\right)$.
The argument generalizes if $L_{n}\left(h_{n}\right)$ is defined
to be the permutation distribution for $T_{n}=\hat{D}_{n,1}\left(\mathbf{X}_{i}\right)$
based on $\lfloor np+\sqrt{n}h_{n}\rfloor$ number of ones, so that
the fixed number of ones at time $n$, $a_{n}$, satisfies $n^{-1/2}\left(a_{n}-np\right)\to h$.

We then generalize this result to derive the limiting permutation
distribution for $T_{n}=\hat{D}_{n,1}\left(\mathbf{X}_{i}\right)$
under stationary alternatives in which the number of ones in $n$
Bernoulli trials converges in distribution under an appropriate normalization.
Note that the permutation distribution $\hat{R}_{n}^{T}$ can be expressed
as $L_{n}\left(\hat{h}_{n}\right)$, where
\[
\hat{h}_{n}=n^{-1/2}\left(\hat{a}_{n}-\frac{n}{2}\right),
\]
and $\hat{a}_{n}$ is the number of ones in $n$ Bernoulli trials.
\begin{thm}
\label{thm: General D_1}Suppose that $\mathbf{X}_{i}=\left\{ X_{ij}\right\} _{j=1}^{\infty}$
is a possibly dependent, stationary Bernoulli sequence. Let $\hat{a}_{n}$
denote the number of ones in the first $n$ elements of $\mathbf{X}_{i}$
and $p_{i}$ denote the marginal probability of a success. Assume
that $n^{-1/2}\left(\hat{a}_{n}-np_{i}\right)$ converges in distribution
to some limiting distribution as $n\to\infty$. Then, the permutation
distribution for $\sqrt{n}T_{n}$ based on the test statistic $T_{n}=\hat{D}_{n,1}\left(X_{i1},\dots,X_{in}\right)$
converges to $N\left(0,1\right)$ in probability; that is
\[
\sup_{t}\vert\hat{R}_{n}^{T}\left(t\right)-\Phi\left(t\right)\vert\overset{P}{\to}0
\]
as $n\to\infty$, where $\Phi\left(\cdot\right)$ denotes the standard
normal cumulative distribution function.
\end{thm}
\begin{rem}
The same argument can be applied to generalize Theorem \ref{thm: General D_1}
for statistics $T_{n}$ equal to $\hat{D}_{n,k}\left(\mathbf{X}_{i}\right)$
or $\hat{P}_{n,k}\left(\mathbf{X}_{i}\right)-\hat{p}_{n,i}$ for general
$k$. $\hfill\blacksquare$
\end{rem}
\begin{cor}
\label{cor: perm alt}Suppose that $n^{-1/2}\left(\hat{a}_{n}-np_{i}\right)$
converges in distribution to some limiting distribution as $n\to\infty$.
Then, if the test statistic $T_{n}$ is equal to $\hat{P}_{n,k}\left(\mathbf{X}_{i}\right)-\hat{p}_{n,i}$
or $\hat{D}_{n,k}\left(\mathbf{X}_{i}\right)$, the permutation distribution
for $\sqrt{n}T_{n}$ satisfies

\[
\sup_{t}\Bigg\vert\hat{R}_{n}^{T}\left(t\right)-\Phi\left(t/\sqrt{\sigma_{T}^{2}\left(p_{i},k\right)}\right)\Bigg\vert\overset{P}{\to}0.
\]
That is, rather than $N\left(0,1\right)$ as the limit, one gets the
same unconditional limiting distribution for these statistics as would
be obtained under i.i.d.\ sampling with success probability $p_{i}$,
where $p_{i}$ denotes the marginal probability of success.
\end{cor}
\begin{rem}
The assumption that $n^{-1/2}\left(\hat{a}_{n}-np_{i}\right)$ converges
in distribution can be weakened to the assumption that $\mathbf{X}_{i}$
is an $\alpha$-mixing process, as the former condition follows from
the latter assumption under stationarity by Theorem 1.7 of \citet{ibragimov1962some}.$\hfill\blacksquare$
\end{rem}
\begin{rem}
In Online Appendix C, we discuss the asymptotic distribution of the
stratified permutation distributions for the test statistics $\sqrt{ns}\bar{P}_{k}\left(\mathbf{X}\right)$
and $\sqrt{ns}\bar{D}_{k}\left(\mathbf{X}\right)$ under the condition
that $n^{-1/2}\left(\hat{a}_{n}-np_{i}\right)$ converges in distribution
for each individual $i$ in $1,\ldots,s$. In particular, we find
that the limiting stratified permutation distribution is the same
unconditional limiting distribution for these statistics as would
be obtained under i.i.d.\ sampling with success probability $p_{i}$
for $i$ in $1,\ldots,s$.$\hfill\blacksquare$
\end{rem}

\section{Power Against a Class of Markov Chain Streaky Alternatives\label{sec:Power}}

In this section, we study the power of the permutation tests developed
in Section \ref{sec:Detecting} against specific models of streaky
alternatives.

\subsection{A Class of Markov Chain Streaky Alternatives\label{subsec:A-Markov-Model}}

We are interested in measuring the power of the permutation tests
developed in Section \ref{sec:Detecting} against stationary alternatives
in which the parameters $\bar{\theta}_{P}^{k}\left(\mathbb{P}\right)$
and $\bar{\theta}_{D}^{k}\left(\mathbb{P}\right)$ are greater than
zero. In this section, we specify a parsimonious class of Markov chain
alternatives of this form. Each instance of these alternatives parameterizes
$\bar{\theta}_{P}^{k}\left(\mathbb{P}\right)$ and $\bar{\theta}_{D}^{k}\left(\mathbb{P}\right)$
with two terms: $\epsilon$ and $\zeta$. The parameter $\epsilon$
determines the ``magnitude'' of the deviation from randomness. The
parameter $\zeta$ determines the ``prevalence'' of the deviation
from randomness across individuals.

There are $s$ total individuals who fall into one of two types --
random and streaky. For each individual, the probability that they
are streaky is $\zeta$. That is, the number of streaky individuals
in the sample is Binomial$(s,\zeta)$. For each individual $i$ in
$1,\ldots,s$, there is a Bernoulli sequence $\mathbf{X}_{i}=\left\{ X_{ij}\right\} _{j=1}^{n}$
of length $n$. Each $\mathbf{X}_{i}$ follows a Markov chain of order
$2^{m}$. The states of the Markov chain are given by the $2^{m}$
binary tuples $\left\{ 0,1\right\} ^{m}$. The event that $\mathbf{X}_{i}$
is in state $(x_{1},\ldots,x_{m})\in\left\{ 0,1\right\} ^{m}$ at
time $j$ corresponds to the event 
\[
X_{ij}=x_{1},X_{i\left(j-1\right)}=x_{2},\ldots,X_{i\left(j-\left(m-1\right)\right)}=x_{m}.
\]

The sequence $\mathbf{X}_{i}$ is i.i.d.\ Bernoulli$(p_{i})$ for
each random individual $i$. That is, for each $(x_{1},\ldots,x_{m})$
in $\left\{ 0,1\right\} ^{m}$, the probabilities of transitioning
to $(1,x_{1},\ldots,x_{m-1})$ and $(0,x_{1},\ldots,x_{m-1})$ are
equal to $p_{i}$ and $\left(1-p_{i}\right)$, respectively. Streaky
individuals deviate from randomness after streaks of $m$ ones or
$m$ zeros. For these individuals the probability of a one or a zero
increases by $\epsilon$ after a streak of $m$ ones or $m$ zeros,
respectively. Formally, for each streaky individual $i$, the probabilities
of transitioning from $\mathbf{1}^{m}$ to itself and $\mathbf{0}^{m}$
to itself are $p_{i}+\epsilon$ and $\left(1-p_{i}\right)+\epsilon$,
where $\mathbf{1}^{m}$ is an $m$-vector of ones, $\mathbf{0}^{m}$
is an $m$-vector of zeros, and $\epsilon$ is a positive real number
less than $\min\left(1-p_{i},p_{i}\right)$. That is, the probabilities
of a one or a zero after a sequence of $m$ ones or $m$ zeros are
equal to $p_{i}+\epsilon$ and $\left(1-p_{i}\right)+\epsilon$, respectively.
For all other states, the transition probabilities are the same as
for a random individual. Observe that, for a streaky individual $i$,
$\theta_{P}^{m}\left(\mathbb{P}_{i}\right)=\epsilon$ and $\theta_{D}^{m}\left(\mathbb{P}_{i}\right)=2\epsilon$.
For a random individual, these parameters are equal to zero. Likewise,
we have that under this model, $\bar{\theta}_{P}^{m}\left(\mathbb{P}\right)=\zeta\epsilon$
and $\bar{\theta}_{D}^{m}\left(\mathbb{P}\right)=2\zeta\epsilon$.

Throughout this section, we specialize to the case that $p_{i}=0.5$
for all individuals. The results are easily generalized to arbitrary
$p_{i}$ with more involved notation. In our empirical setting of
controlled basketball shooting experiments, shot locations are chosen
such that shooting percentages should be close to $0.5$. The average
shooting percentages for the GVT and \citet{miller2018cold} shooting
experiments are 52\% and 50\%, respectively.

\subsection{Analytic Power Approximation\label{subsec:Analytic-Power-Approximation}}

In this subsection, we derive analytic asymptotic approximations to
the power of the permutation tests presented in Section \ref{thm: Convergence_under_H0}
against the class of streaky alternatives specified in Section \ref{subsec:A-Markov-Model}.
For the sake of parsimony, we present the details of our argument
when $m=k=1$ and conclude with a discussion of the generalization
of these results to cases with $m$ and $k$ greater than one. The
details of this generalization appear in Online Appendix H.

First, we characterize the exact asymptotic distribution of $\hat{P}_{n,1}\left(\mathbf{X}_{i}\right)-\hat{p}_{n,i}$
and $\hat{D}_{n,1}\left(\mathbf{X}_{i}\right)$ computed on the Bernoulli
sequence $\mathbf{X}_{i}$ of a streaky individual. By invoking the
characterizations of the limiting permutation distributions developed
in Section \ref{subsec:Asymptotic-Permutation}, we can compute the
limiting power of the permutation tests that we study against streaky
alternatives local to the null hypothesis.
\begin{thm}
\label{thm: Alternative m =00003D 1}Assume that $\mathbf{X}_{i}=\left\{ X_{ij}\right\} _{j=1}^{n}$
is a two-state stationary Markov Chain on $\left\{ 0,1\right\} $
with transition matrix given by 
\begin{equation}
\mathcal{P}=\left[\begin{array}{cc}
\frac{1}{2}+\epsilon & \frac{1}{2}-\epsilon\\
\frac{1}{2}-\epsilon & \frac{1}{2}+\epsilon
\end{array}\right],\label{eq: transition}
\end{equation}
where $0\leq\epsilon<\frac{1}{2}$. Then:

\noindent \textbf{(i)} $\hat{P}_{n,1}\left(\mathbf{X}_{i}\right)-\hat{p}_{n,i}$,
with $\hat{P}_{n,1}\left(\mathbf{X}_{i}\right)$ given by (\ref{eq:P_k})
with $k$ equal to $1$ and $\hat{p}_{n,i}=n^{-1}\sum_{j=1}^{n}X_{ij}$,
is asymptotically normal with limiting distribution given by

\noindent 
\[
\sqrt{n}\left(\hat{P}_{n,1}\left(\mathbf{X}_{i}\right)-\hat{p}_{n,i}-\epsilon\right)\overset{d}{\rightarrow}N\left(0,\frac{1-2\epsilon+16\epsilon^{2}}{4-8\epsilon}\right),
\]
and

\noindent \textbf{(ii)} $\hat{D}_{n,1}\left(\mathbf{X}_{i}\right)$,
given by (\ref{eq:D_k}) with $k$ equal to $1$, is asymptotically
normal with limiting distribution given by

\noindent 
\[
\sqrt{n}\left(\hat{D}_{n,1}\left(\mathbf{X}_{i}\right)-2\epsilon\right)\overset{d}{\rightarrow}N\left(0,1-4\epsilon^{2}\right)
\]
as $n\to0$.
\end{thm}
\begin{rem}
The argument for Theorem \ref{thm: Alternative m =00003D 1} holds
if we let $\epsilon$ vary with $n$ such that $\epsilon_{n}=\epsilon+O\left(n^{-1/2}\right)$.
In particular, if we take $\epsilon_{n}=\frac{h}{\sqrt{n}},$ then
\begin{eqnarray*}
\sqrt{n}\left(\hat{P}_{n,1}\left(\mathbf{X}_{i}\right)-\hat{p}_{n,i}\right)\overset{d}{\to}N\left(h,1/4\right) & \text{ and } & \sqrt{n}\hat{D}_{n,1}\left(\mathbf{X}\right)\overset{d}{\to}N\left(2h,1\right).
\end{eqnarray*}
Additionally, under the conditions of Theorem \ref{thm: Alternative m =00003D 1},
Remark 3.7 indicates that as $\mathbf{X}_{i}=\left\{ X_{ij}\right\} _{j=1}^{n}$
is $\alpha$-mixing, $n^{-1/2}\left(\hat{a}_{n}-np_{i}\right)$ converges
in distribution to some limiting distribution as $n\to\infty$, where
$\hat{a}_{n}$ denotes the number of ones in the first $n$ elements
of $\mathbf{X}_{i}$. Thus, by Corollary \ref{cor: perm alt} and
Lemma 11.2.1 of \citet{lehmann2005testing}, the $1-\alpha$ quantile
of the permutation distribution for $\hat{D}_{n,1}\left(\mathbf{X}\right)$
converges in probability to $z_{1-\alpha}$ -- the $1-\alpha$ quantile
of the standard normal distribution. Hence, by Slutsky's Theorem,
the power of the permutation test with test statistic $T_{n}$ equal
to $\hat{D}_{n,1}\left(\mathbf{X}_{i}\right)$ is given by 
\begin{flalign*}
\mathbb{P}_{i}\left\{ \sqrt{n}\hat{D}_{n,1}\left(\mathbf{X}_{i}\right)>\hat{r}_{n}^{\hat{D}_{1}}\left(1-\alpha\right)\right\}  & =\mathbb{P}_{i}\left\{ \sqrt{n}\left(\hat{D}_{n,1}\left(\mathbf{X}_{i}\right)-\frac{2h}{\sqrt{n}}\right)>\hat{r}_{n}^{\hat{D}_{1}}\left(1-\alpha\right)-2h\right\} \\
 & \to1-\Phi\left(z_{1-\alpha}-2h\right)
\end{flalign*}
as $n\to\infty$, where $\hat{r}_{n}^{\hat{D}_{1}}\left(1-\alpha\right)$
denotes the $1-\alpha$ quantile of the permutation distribution of
$\hat{D}_{n,1}\left(\mathbf{X}_{i}\right)$. An analogous result holds
for the permutation tests with test statistic $T_{n}$ equal to $\sqrt{n}\left(\hat{P}_{n,1}\left(\mathbf{X}_{i}\right)-\hat{p}_{n,i}\right)$.
This argument implies the following Corollary.$\hfill\blacksquare$
\end{rem}
\begin{cor}
\label{cor: indiv power m 1} Consider the permutation test of the
null hypothesis $H_{0}^{i}$ that the Bernoulli sequence $\mathbf{X}_{i}=\left\{ X_{ij}\right\} _{j=1}^{n}$
is i.i.d.\ rejecting for large values of the test statistic $T_{n}$.
If the test statistic $T_{n}$ is equal to $\hat{P}_{n,1}\left(\mathbf{X}_{i}\right)-\hat{p}_{n,i}$
or $\hat{D}_{n,1}\left(\mathbf{X}_{i}\right)$, then the power of
this test against the alternative that $\mathbf{X}_{i}$ is a two-state
Markov Chain on $\left\{ 0,1\right\} $ with transition matrix given
by (\ref{eq: transition}) and $\epsilon=h/\sqrt{n}$ converges to
$1-\Phi\left(z_{1-\alpha}-2h\right)$ as $n\to\infty$ .
\end{cor}
Moreover, Theorem \ref{thm: Alternative m =00003D 1} allows us to
characterize the limiting distributions of $\bar{P}_{k}\left(\mathbf{X}\right)$
and $\bar{D}_{k}\left(\mathbf{X}\right)$ under the Markov chain streaky
alternatives. In turn, this result allows us to derive an expression
for the limiting power of stratified permutation tests of the joint
null hypothesis $H_{0}$ against streaky alternatives local to the
joint null hypothesis that use $\bar{P}_{k}\left(\mathbf{X}\right)$
and $\bar{D}_{k}\left(\mathbf{X}\right)$ as test statistics.
\begin{cor}
\label{cor: joint power m 1}Assume that a population of $s$ individuals
are associated with the two-state stationary Markov chains $\mathbf{X}_{i}=\left\{ X_{ij}\right\} _{j=1}^{\infty}$
on $\left\{ 0,1\right\} $ for each $i$ in $1,\ldots,s$, such that
each sequence $\mathbf{X}_{i}$ has probability $\zeta$ of having
transition matrix given by (\ref{eq: transition}) with $\epsilon=h/\sqrt{ns}$
and is otherwise i.i.d.\ Bernoulli$(1/2)$. Then:

\noindent \textbf{(i)} $\bar{P}_{1}\left(\mathbf{X}\right)$, given
by (\ref{eq: averages}) with $k=1$, is asymptotically normal with
limiting distribution given by 
\[
\sqrt{ns}\bar{P}_{1}\left(\mathbf{X}\right)\overset{d}{\to}N\left(\zeta h,1/4\right),
\]
and

\noindent \textbf{(ii)} $\bar{D}_{1}\left(\mathbf{X}\right)$, given
by (\ref{eq: averages}) with $k=1$, is asymptotically normal with
limiting distribution given by 
\[
\sqrt{ns}\bar{D}_{1}\left(\mathbf{X}\right)\overset{d}{\to}N\left(2\zeta h,1\right)
\]
as $n\to\infty$ and $s\to\infty$.

\noindent \textbf{(iii)} The power of the stratified permutation test
of the joint null hypothesis $H_{0}$ rejecting for large values of
the test statistic $K_{n,s}$, for $K_{n,s}$ equal to $\bar{P}_{1}\left(\mathbf{X}\right)$
or $\bar{D}_{1}\left(\mathbf{X}\right)$, against the alternative
specified in the conditions of this corollary converges to $1-\Phi\left(z_{1-\alpha}-2\zeta h\right)$
as $n\to\infty$ and $s\to\infty$.
\end{cor}
Now, we discuss the extension of these results to cases with general
$m$ and $k$. Details of these extensions are given in Online Appendix
H. Consider the Bernoulli sequence of a single individual $\mathbf{X}_{i}$.
The power of the permutation test of the null hypothesis $H_{0}^{i}$
that individual $i$ is random against the alternative that individual
$i$ is streaky with $\epsilon=h/\sqrt{n}$, rejecting for large values
of the test statistic, $T_{n}$ converges to 
\[
1-\Phi\left(z_{1-\alpha}-\phi_{T}\left(k,m,h\right)\right)
\]
for $T$ equal to $P$ or $D$ when $T_{n}$ is equal to $\hat{P}_{n,k}\left(\mathbf{X}_{i}\right)-\hat{p}_{n,i}$
or $\hat{D}_{n,k}\left(\mathbf{X}_{i}\right)$, respectively. The
constant $\phi_{T}\left(k,m,h\right)$ is a function of $k$, $m$,
and $h$ and is given by 
\[
\phi_{T}\left(\text{\ensuremath{k,}}m,h\right)=\lim_{n\to\infty}\frac{\text{\ensuremath{\sqrt{n}}\ensuremath{\mu_{T}\left(k,m,\epsilon_{n}\right)}}}{\sqrt{\sigma_{T}^{2}\left(1/2,k\right)}}
\]
where $\mu_{P}\left(k,m,\epsilon_{n}\right)$ and $\mu_{D}\left(k,m,\epsilon_{n}\right)$
are the asymptotic means of $\hat{P}_{n,k}\left(\mathbf{X}_{i}\right)-\hat{p}_{n,i}$
or $\hat{D}_{n,k}\left(\mathbf{X}_{i}\right)$ when $\mathbf{X}_{i}$
corresponds to a streaky individual with $\epsilon_{n}=h/\sqrt{n}$,
and $\sigma_{P}^{2}\left(1/2,k\right)$ and $\sigma_{D}^{2}\left(1/2,k\right)$
are the asymptotic variances of $\hat{D}_{n,k}\left(\mathbf{X}_{i}\right)$
under $H_{0}$, given by Theorem \ref{thm: Convergence_under_H0}.
We give expressions for $\mu_{P}\left(k,m,\epsilon\right)$ and $\mu_{D}\left(k,m,\epsilon\right)$
in terms of $k$, $m$, and $\epsilon$ in Online Appendix H. Corollary
\ref{cor: indiv power m 1} shows that in the case that $m$ and $k$
are equal to one, $\phi_{T}\left(k,m,h\right)=2h$ for both $T$ equal
to $D$ and $P$.

Table \ref{tab:asym power} displays the values of $\phi_{D}\left(\text{\ensuremath{k,}}m,h\right)$
for $m$ and $k$ between one and four. The permutation tests that
reject for large values of $\hat{D}_{n,k}\left(\mathbf{X}_{i}\right)$
with $k$ equal to $m$ have the largest power against the alternative
where deviations from randomness begin after $m$ consecutive ones
or zeros. The permutation test that rejects for large values $\hat{D}_{n,k}\left(\mathbf{X}\right)$
with $k$ equal to one against the alternative with $m$ equal to
one has the largest power over any combination of the test statistics
and alternatives that we consider. Thus, the power of the test using
$\hat{D}_{n,k}\left(\mathbf{X}_{i}\right)$ with $k$ equal to one
against the alternative with $m$ equal to one gives an upper bound
to the power of any of the permutation tests that we consider against
any of the Markov chain streaky alternatives for a given value of
$\epsilon$. In fact, in Online Appendix I, we show that the permutation
tests rejecting for large values of $\hat{D}_{n,k}\left(\mathbf{X}_{i}\right)$
and $\hat{P}_{n,k}\left(\mathbf{X}_{i}\right)-\hat{p}_{n,i}$ for
$k$ equal to one are asymptotically equivalent to the uniformly most
powerful unbiased test against first order Markov chains.

\begin{table}[t]
\begin{centering}
\begin{tabular}{cccccc}
\toprule 
\multicolumn{6}{c}{$m$}\tabularnewline
\midrule 
$k$ &  & 1 & 2 & 3 & 4\tabularnewline
\midrule
\midrule 
1 &  & $2h$ & $h$ & $\frac{h}{2}$ & $\frac{h}{4}$\tabularnewline
\midrule 
2 &  & $\sqrt{2}h$ & $\sqrt{2}h$ & $\frac{h}{\sqrt{2}}$ & $\frac{h}{2\sqrt{2}}$\tabularnewline
\midrule 
3 &  & $h$ & $h$ & $h$ & $\frac{h}{2}$\tabularnewline
\midrule 
4 &  & $\frac{h}{\sqrt{2}}$ & $\frac{h}{\sqrt{2}}$ & $\frac{h}{\sqrt{2}}$ & $\frac{h}{\sqrt{2}}$\tabularnewline
\bottomrule
\end{tabular}
\par\end{centering}
\medskip{}

\begin{centering}
\caption{\label{tab:asym power}Value of $\phi_{D}\left(\text{\ensuremath{k,}}m,h\right)$
For Small Values of $k$ and $m$}
\par\end{centering}
\medskip{}

{\footnotesize{}Notes: Table displays the limit as $n$ grows to infinity
of the $\sqrt{n}$ scaled ratio of the mean and standard deviation
of the asymptotic distribution of $\hat{D}_{n,k}\left(\mathbf{X}_{i}\right)$
under the Markov chain alternatives considered in Section \ref{subsec:A-Markov-Model}
for $m$ and $k$ between one and four with local perturbations $\epsilon_{n}=\frac{h}{\sqrt{n}}$.}{\footnotesize\par}
\end{table}

Similarly, consider a collection of Bernoulli sequences for $s$ individuals.
The power of the stratified permutation test of the joint null hypothesis
$H_{0}$ -- that all of the individuals are random -- against the
alternative that each individual is streaky with probability $\zeta$
and $\epsilon=h/\sqrt{ns}$ rejecting for large values of the test
statistic $K_{n}^{T}$ converges to
\[
1-\Phi\left(z_{1-\alpha}-\phi_{T}\left(k,m,h\right)\zeta\right)
\]
for $T$ equal to $P$ or $D$ when the test statistic $K_{n,s}$
is equal to $\bar{P}_{k}\left(\mathbf{X}\right)$ or $\bar{D}_{k}\left(\mathbf{X}\right)$,
respectively.

\begin{figure}[t]
\caption{\label{fig: Necessary Observations}Requisite Sample Size for Power
of Tests of the Joint Null Against Specified Alternatives}

\medskip{}

\begin{centering}
\begin{tabular}{c}
\textit{\small{}\includegraphics[scale=0.31]{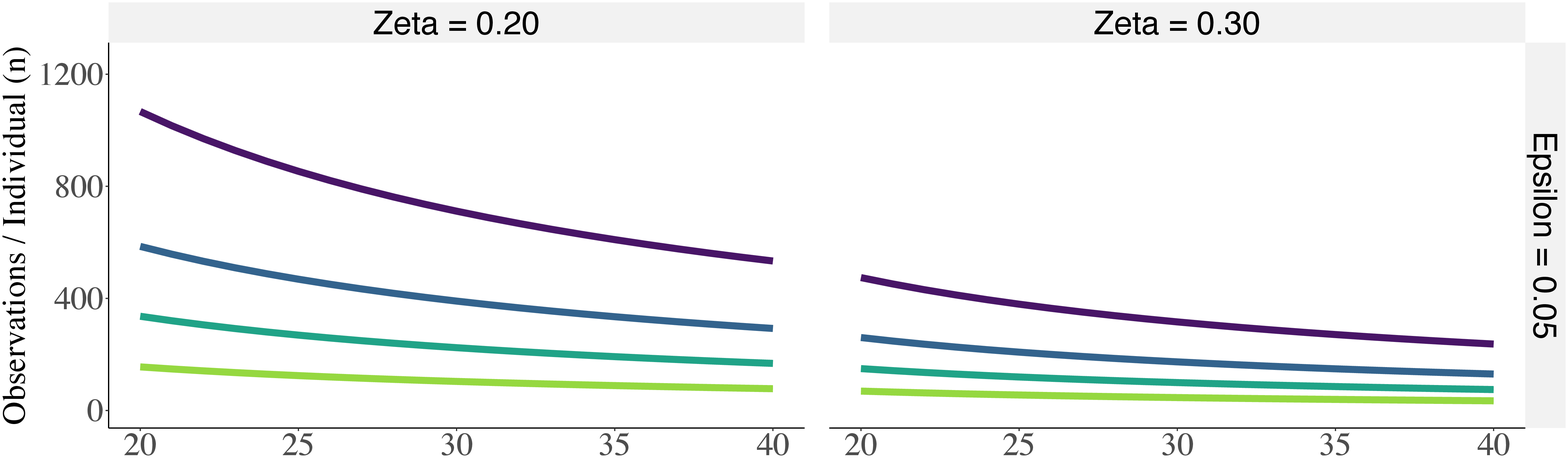}}\tabularnewline
\textit{\small{}\includegraphics[scale=0.31]{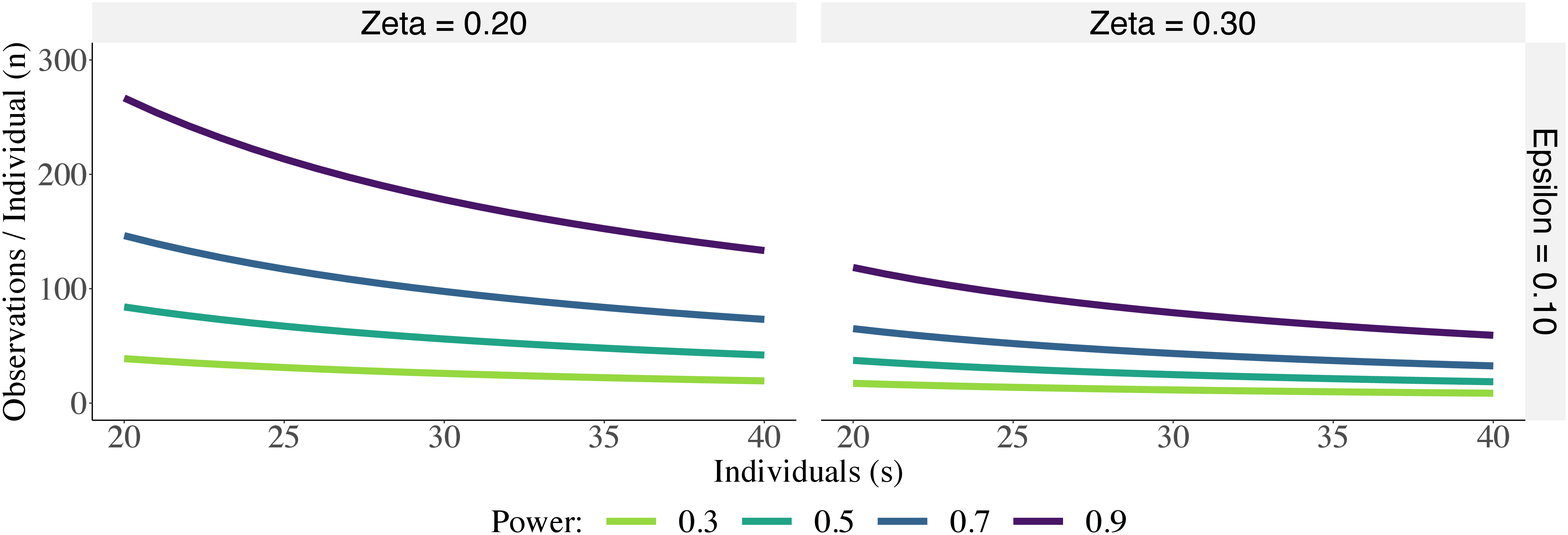}}\tabularnewline
\end{tabular}
\par\end{centering}
\medskip{}

{\footnotesize{}Notes: Figure displays the power of the permutation
test of the joint null hypothesis $H_{0}$ using the test statistic
$\bar{D}_{1}\left(\mathbf{X}\right)$ against the Markov chain streaky
alternative with $m=1$ calculated by the analytic approximation in
Corollary \ref{cor: joint power m 1}. Each panel gives the power
for the test for different sample sizes $n$ and $s$ under a specified
$\epsilon$ and $\zeta$.}{\footnotesize\par}
\end{figure}

These results are very useful for power calculations when planning
or assessing experiments. Suppose that we were planning on implementing
an experiment where we would collect Bernoulli sequences $\mathbf{X}_{i}$
of length $n$ from $s$ individuals, and would like to test the joint
null hypothesis $H_{0}$ that all sequences are i.i.d.\ against an
alternative where each sequence is streaky with probability $\zeta$
for a given $\epsilon$ and $m$ equal to one with a desired power
of $\beta$. If we use the test statistic $T_{n}$ equal to $\bar{D}_{1}$,
our results demonstrate that the product of the number of individuals
$s$ and observations per individual $n$ should be approximately
\begin{equation}
\left(\frac{z_{1-\alpha}-z_{1-\beta}}{2\zeta\epsilon}\right)^{2}.\label{eq: power D_1}
\end{equation}
This calculation is straightforward for any choice of parameter values,
test statistic, and $m$ by plugging in $h=\epsilon\sqrt{ns}$ and
solving for $ns$ in the limiting power expression from Corollary
\ref{cor: indiv power m 1}. Figure \ref{fig: Necessary Observations}
displays the power of the test rejecting for large values of $\bar{D}_{1}\left(\mathbf{X}\right)$
at level $\alpha=0.05$ against four parameterizations of $\epsilon$
and $\zeta$ for the Markov chain streaky alternative with $m=1$
for a grid of values of $n$ and $s$. As we outline in the subsequent
section, measuring these power curves with simulation is computationally
very costly.

\subsection{Simulation Analysis\label{subsec:Simulation-Analysis}}

In this section, we study the finite-sample quality of the asymptotic
approximations to the power of the permutation tests that we consider
against the Markov chain streaky alternative specified in Section
\ref{subsec:A-Markov-Model}. We focus on permutation tests of the
individual hypotheses $H_{0}^{i}$ that use the test statistic $\hat{D}_{n,k}\left(\mathbf{X}_{i}\right)$
and of the joint hypothesis $H_{0}$ that use the test statistic $\bar{D}_{k}\left(\mathbf{X}\right)$.
The results for permutation tests using $\hat{P}_{n,k}\left(\mathbf{X}_{i}\right)-\hat{p}_{n,i}$
and $\bar{P}_{k}\left(\mathbf{X}\right)$ are very similar.

The simulations presented in this section require extensive parallelization.
The computation is particularly expensive for measurement of the power
of tests of the joint hypothesis $H_{0}$, as the permutation distributions
of joint test statistics of each draw of $s$ individuals need to
be computed.\footnote{The simulation underlying Figure \ref{fig:Contours} utilizes 2,600
nodes, each equipped with 15 cores. If the script were run in serial,
it would take approximately five years and six months to run to completion.} In contrast, measuring the minimum $n$ and $s$ necessary to achieve
a desired power against a wide range of $\epsilon$ and $\zeta$ is
instantaneous with the analytic approximation given in Corollary \ref{cor: joint power m 1}.

\begin{figure}[t]
\caption{\label{fig:finite power curve}Power Curve for Permutation Test Rejecting
for Large $\hat{D}_{n,k}\left(\mathbf{X}_{i}\right)$}

\medskip{}

\begin{centering}
\begin{tabular}{c||c}
\multicolumn{2}{c}{\textit{\includegraphics[scale=0.4]{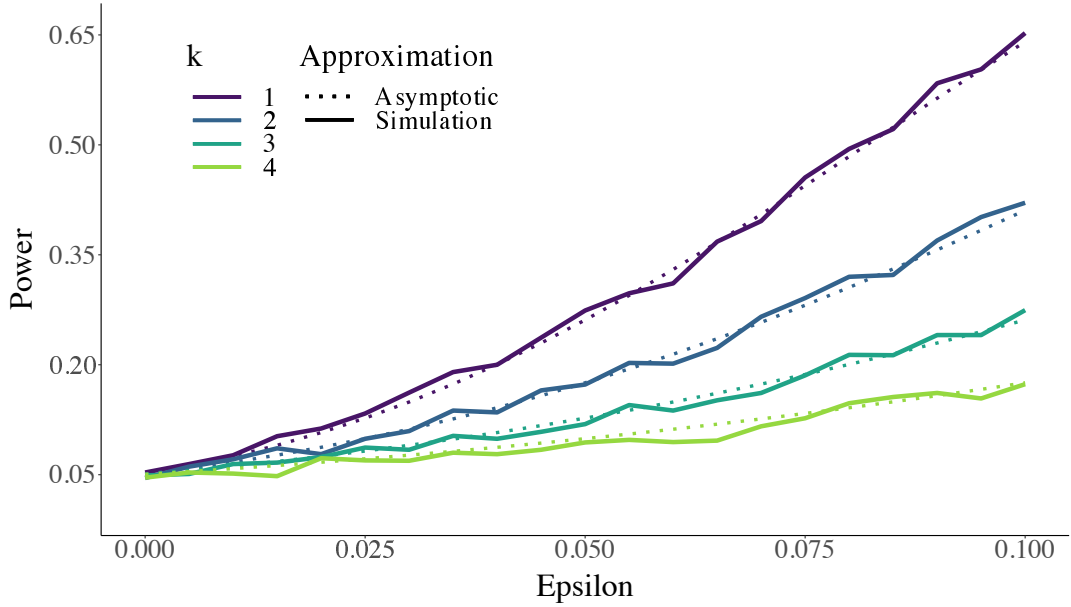}}}\tabularnewline
\end{tabular}
\par\end{centering}
\medskip{}

{\footnotesize{}Notes: Figure displays the power for the permutation
test rejecting at level $\alpha=0.05$ for large values of $\hat{D}_{n,k}\left(\mathbf{X}_{i}\right)$
for a range of $\epsilon$ in the alternative given by (\ref{eq: transition}),
$n=100$, and each $k$ in $1,\ldots,4$. The solid lines display
the power measured by a simulation, which takes the proportion of
$2,000$ replications of Bernoulli sequences $\mathbf{X}_{i}$ following
the transition matrix (\ref{eq: transition}) on which the permutation
test using $\hat{D}_{n,k}\left(\mathbf{X}_{i}\right)$ rejects $H_{0}^{i}$
at $5\%$ level for each value of $\epsilon$. The dashed lines display
the power calculated by the analytic approximation given by Corollary
\ref{cor: indiv power m 1}.}{\footnotesize\par}
\end{figure}

Figure \ref{fig:finite power curve} displays the power for the permutation
test that rejects at level $0.05$ for large values of $\hat{D}_{n,k}\left(\mathbf{X}_{i}\right)$
for $k$ between $1$ and $4$ and $n$ equal to $100$ against the
alternative that $\mathbf{X}_{i}$ is a Bernoulli sequence associated
with a streaky individual with $m$ equal to one over a grid of $\epsilon$.\footnote{Most shooters take 100 shots in the experiment considered in GVT and
MS. Three shooters take 90, 75, and 50 shots, respectively.} The solid lines display the power of each test measured with a simulation,
drawing and implementing the tests on 2,000 replicates of sequences
for each value of $\epsilon$. The dashed lines display the power
approximated with the asymptotic expression given in Corollary \ref{cor: indiv power m 1}.
The finite-sample simulation and asymptotic approximations are remarkably
close.

\begin{figure}[t]
\caption{\label{fig:Contours}Power Contours for Permutation Test Rejecting
for Large $\bar{D}_{1}\left(\mathbf{X}\right)$}

\medskip{}

\begin{centering}
\begin{tabular}{c}
\textit{\small{}\includegraphics[scale=0.4]{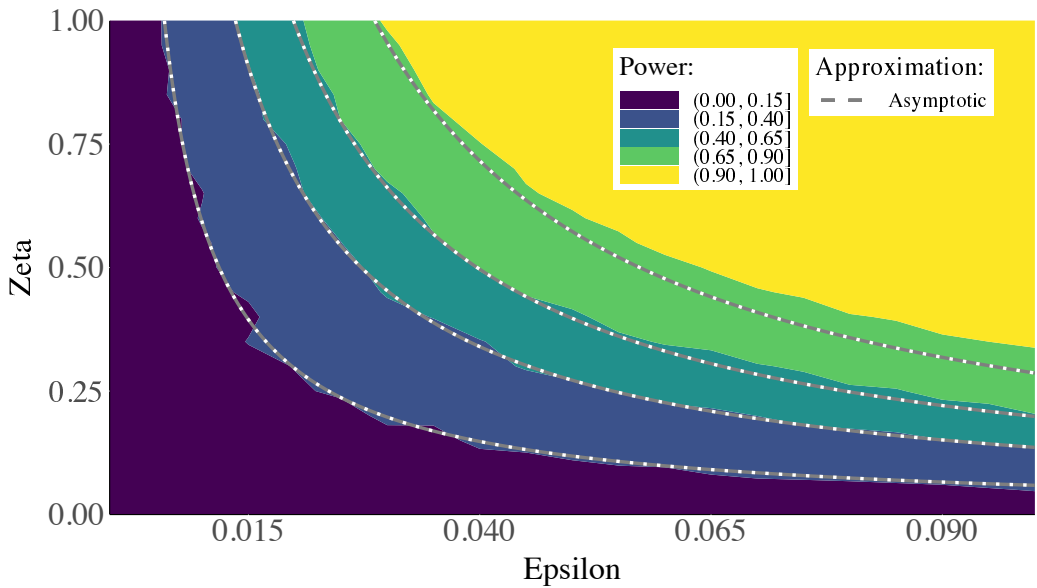}}\tabularnewline
\end{tabular}
\par\end{centering}
\medskip{}

{\footnotesize{}Notes: Figure displays contours of the power surface
on $\epsilon$ and $\zeta$ for the stratified permutation test rejecting
at level $0.05$ for large values of $\bar{D}_{1}\left(\mathbf{X}\right)$
against the streaky alternative specified in Section \ref{subsec:A-Markov-Model}
for $n$ equal to $100$, $s$ equal to $26$, and $m=1$. We draw
1,000 replicates of $s$ Bernoulli sequences $\mathbf{X}_{i}$ according
to the streaky alternative specified in Section \ref{subsec:A-Markov-Model}
with $m=1$ for each $\epsilon$ and $\zeta.$ The estimate of the
power at each $\epsilon$ and $\zeta$ is given by the proportion
of replicates in which the stratified permutation test using $\bar{D}_{1}\left(\mathbf{X}\right)$
rejects $H_{0}$ at level $0.05$. The estimates of power are grouped
into five regions, corresponding to sets of $\epsilon$ and $\zeta$
values with estimated power in five mutually exclusive intervals on
$(0,1]$. The white dotted curves give the asymptotic approximations
to the $\zeta$ values at which the permutation test rejecting for
large values of $\bar{D}_{1}\left(\mathbf{X}\right)$ at level $0.05$
has power equal to $0.15$, $0.40$, $0.65$, $0.90$, and $1.00$
as a function of $\epsilon$. The expressions for these curves are
obtained by solving expression (\ref{eq: power D_1}) for $\zeta$
for a given value of $\beta$ in terms of $\epsilon$.}{\footnotesize\par}
\end{figure}

Figure \ref{fig:Contours} displays contours of the power surface
on $\epsilon$ and $\zeta$ for the stratified permutation test rejecting
at level $0.05$ for large values of $\bar{D}_{1}\left(\mathbf{X}\right)$
against the streaky alternative specified in Section \ref{subsec:A-Markov-Model}
for $n$ equal to $100$, $s$ equal to $26$, and $m=1$.\footnote{There are 26 individuals who participate in the GVT controlled shooting
experiment. For all but three individuals, we observe 100 shots. We
simulate 100 shots for each individual and so compute a slight upper
bound to the power of the tests that we consider.} For each $\epsilon$ and $\zeta$ on a two dimensional grid, we measure
the power of the permutation test rejecting for large $\bar{D}_{1}\left(\mathbf{X}\right)$
with simulation by drawing and implementing the test on 1,000 replicates
of $s$ sequences. We find that our asymptotic approximation is very
accurate for most parameterizations of the model at the sample size
that we study. However, our approximation appears to overestimate
the power for parameterizations where the finite-sample power is close
to $0.9$.

\section{Uncertainty in the Hot Hand Fallacy\label{sec:hot-hand}}

In the preceding two sections, we developed inferential methods for
testing whether a collection of Bernoulli sequences deviates from
randomness. Equipped with these methods, we now examine two empirical
questions posed formally in Section \ref{sec:Posing-the-Problem}.
First, is there evidence of positive serial dependence in basketball
shooting? Second, if so, how widespread and substantial is this dependence?

We begin by providing an overview of the available data from controlled
basketball shooting experiments, before addressing these two questions
in succession. We conclude by discussing evidence on beliefs in serial
dependence in basketball shooting, outlining a formal framework for
addressing the final, behaviorally substantive, question from Section
\ref{sec:Posing-the-Problem} -- whether people systematically overestimate
positive serial dependence in basketball shooting.

\subsection{Controlled Shooting Experiments}

We examine the evidence for serial dependence in basketball shooting
provided by controlled shooting experiments. In a controlled shooting
experiment, each individual is observed taking a sequence of shots
under identical conditions. Although live game data from professional
and collegiate basketball is abundant, these data are subject to large
and ambiguous selection biases. In a live game setting, making a shot
may subsequently affect defensive pressure, shot selection, and offensive
strategy. Controlling for these effects is a complicated computational
and statistical problem \citep{bocskocsky2014hot,lantis2019hot}.
Controlled shooting experiments provide a significantly cleaner statistical
setting.

We consider the design and results of four controlled shooting experiments.
The GVT shooting experiment is the only experiment designed for tests
for serially dependent shooting whose data are publicly available
and whose results have been peer-reviewed. Moreover, the conclusions
reached in GVT and MS based on the data from the GVT shooting experiment
are starkly different and have resulted in both the former consensus
and current uncertainty concerning the empirical support for the hot
hand fallacy in economics. Thus, we focus on the results of this experiment.

In the GVT shooting experiment, we observe shooting sequences for
26 members of the Cornell University men and women's varsity and junior
varsity basketball teams.\footnote{We obtained the data from \citet{miller2018surprisedsupp}, available
at https://www.econometricsociety.org/sites/default/files/14943\_Data\_and\_Programs.zip
on April 19, 2019.} Fourteen of the players are men and twelve of the players are women.
For all but three players, we observe 100 shots. We observe 90, 75,
and 50 shots for three of the men. The experimenters determined distances
from the basket at which each player's shooting percentage was approximately
50\% and placed two arcs 60 degrees from the baseline on the left
and right hand sides of the basket. Each individual took 50\% of their
shots from each side of the basket. The experiment was incentivized.

\citet{miller2018cold} and \citet{miller2019fallacy} study the results
of three additional controlled shooting experiments. \citet{miller2018cold}
implement an experiment with ten semi-professional Spanish basketball
players. Two shooters took 300 consecutive shots in one session, seven
shooters took 300 consecutive shots in each of three sessions, and
one shooter took 300 shots in each of five sessions. \citet{miller2018cold}
also study data from the controlled shooting experiment presented
originally in \citet{Jagacinski1979}, in which six former collegiate
players took 60 shots in each of nine sessions. The implementations
of these experiments are otherwise very similar to the GVT experiment.
\citet{miller2019fallacy} study the results from the annual NBA Three
Point Shooting contest, in which players compete by taking rounds
of 25 consecutive three point shots. They consider all 34 players
who have taken more than 100 shots in this contest over the course
of their careers. The average number of shots taken in this sample
is 166.

\subsection{Is There Positive Serial Dependence in Basketball Shooting?}

\begin{table}[t]
\begin{centering}
\begin{tabular}{ccccccc}
\toprule 
 &  & \multicolumn{2}{c}{Stratified Permutation Test of $H_{0}$ $p$-Value} &  & \multicolumn{2}{c}{Number of Simultaneous Rejections of $H_{0}^{i}$}\tabularnewline
\cmidrule{3-4} \cmidrule{4-4} \cmidrule{6-7} \cmidrule{7-7} 
k &  & $\bar{P}_{k}\left(\mathbf{X}\right)$ & $\bar{D}_{k}\left(\mathbf{X}\right)$ &  & $\hat{P}_{n,k}(\mathbf{X}_{i})-\hat{p}_{n,i}$ & $\hat{D}_{n,k}(\mathbf{X}_{i})$\tabularnewline
\cmidrule{1-1} \cmidrule{3-4} \cmidrule{4-4} \cmidrule{6-7} \cmidrule{7-7} 
 &  & (1) & (2) &  & (3) & (4)\tabularnewline
\midrule 
1 &  & 0.155 & 0.146 &  & 1 & 1\tabularnewline
\midrule
2 &  & 0.032 & 0.040 &  & 1 & 2\tabularnewline
\midrule
3 &  & 0.042 & 0.004 &  & 1 & 1\tabularnewline
\midrule
4 &  & 0.303 & 0.072 &  & 0 & 0\tabularnewline
\bottomrule
\end{tabular}
\par\end{centering}
{\small{}\medskip{}
}{\small\par}

\caption{\label{tab:Results}Results of Simultaneous and Joint Hypothesis Tests
for the GVT Experiment}

\medskip{}

{\footnotesize{}Notes: Table displays the results of simultaneous
tests of the individual null hypotheses $H_{0}^{i}$ and tests of
the joint null hypothesis $H_{0}$ for the GVT controlled shooting
experiment. Columns (1) and (2) display the $p$-values of the stratified
permutation test of $H_{0}$ using the statistics $\bar{P}_{k}\left(\mathbf{X}\right)$
and $\bar{D}_{k}\left(\mathbf{X}\right)$, respectively. We estimate
the stratified permutation distribution of each statistic with 100,000
stratified permutations. Columns (3) and (4) display the number of
rejections of $H_{0}^{i}$ at level $\alpha=0.05$ using the test
statistics $\hat{D}_{n,k}(\mathbf{X}_{i})$ and $\hat{P}_{n,k}(\mathbf{X}_{i})-\hat{p}_{n,i}$
for each $k$ in $1,\ldots,4$, respectively. We use the stepdown
procedure with \v{S}id\'{a}k critical values implemented on the
$p$-values from the one-sided permutation test.}{\footnotesize\par}
\end{table}

The first question posed in Section \ref{sec:Posing-the-Problem}--
whether all basketball shooting is random -- can be assessed with
a test of the joint null hypothesis $H_{0}$. Columns (1) and (2)
of Table \ref{tab:Results} display the $p$-values for the stratified
permutation tests of $H_{0}$ using $\bar{P}_{k}\left(\mathbf{X}\right)$
and $\bar{D}_{k}\left(\mathbf{X}\right)$ in the GVT shooting experiment,
respectively. Both tests reject at the 5\% level for $k$ equal to
$2$ and $3$. The test using $\bar{D}_{k}\left(\mathbf{X}\right)$
for $k$ equal to $3$ rejects at the $1$\% level. These results
provide reasonably strong evidence that basketball shooting is not
random. However, one may be concerned that the rejection of $H_{0}$
is not overwhelmingly strong.

Assuaging this concern, columns (3) and (4) of Table \ref{tab:Results}
display the number of rejections of $H_{0}^{i}$ at level $\alpha=0.05$
when the $p$-values from the individual shooter permutation tests
using $\hat{P}_{n,k}(\mathbf{X}_{i})-\hat{p}_{n,i}$ and $\hat{D}_{n,k}(\mathbf{X}_{i})$
are corrected with the stepdown procedure with \v{S}id\'{a}k critical
values. The results are identical at level $\alpha=0.1$. The procedure
consistently rejects $H_{0}^{i}$ for only one shooter, identified
as ``Shooter 109,'' over the set of test statistics considered.\footnote{Tables giving the $p$-values of the individual permutation tests
using $\hat{P}_{n,k}(\mathbf{X}_{i})-\hat{p}_{n,i}$ and $\hat{D}_{n,k}(\mathbf{X}_{i})$
for $k$ in $1,\ldots,4$ for each shooter in the GVT shooting experiment
are given in Online Appendix J.}

\begin{figure}[t]
\caption{\label{fig:109}Shooter 109 Shooting Sequence and Permutation Distribution}

\medskip{}

\begin{centering}
\begin{tabular}{cc}
\textit{\small{}Panel A: Shooting Sequence} & \textit{\small{}Panel B: Permutation Distribution of $\hat{D}_{n,1}\left(\mathbf{X}_{i}\right)$}\tabularnewline
\textit{\small{}\includegraphics[scale=0.355]{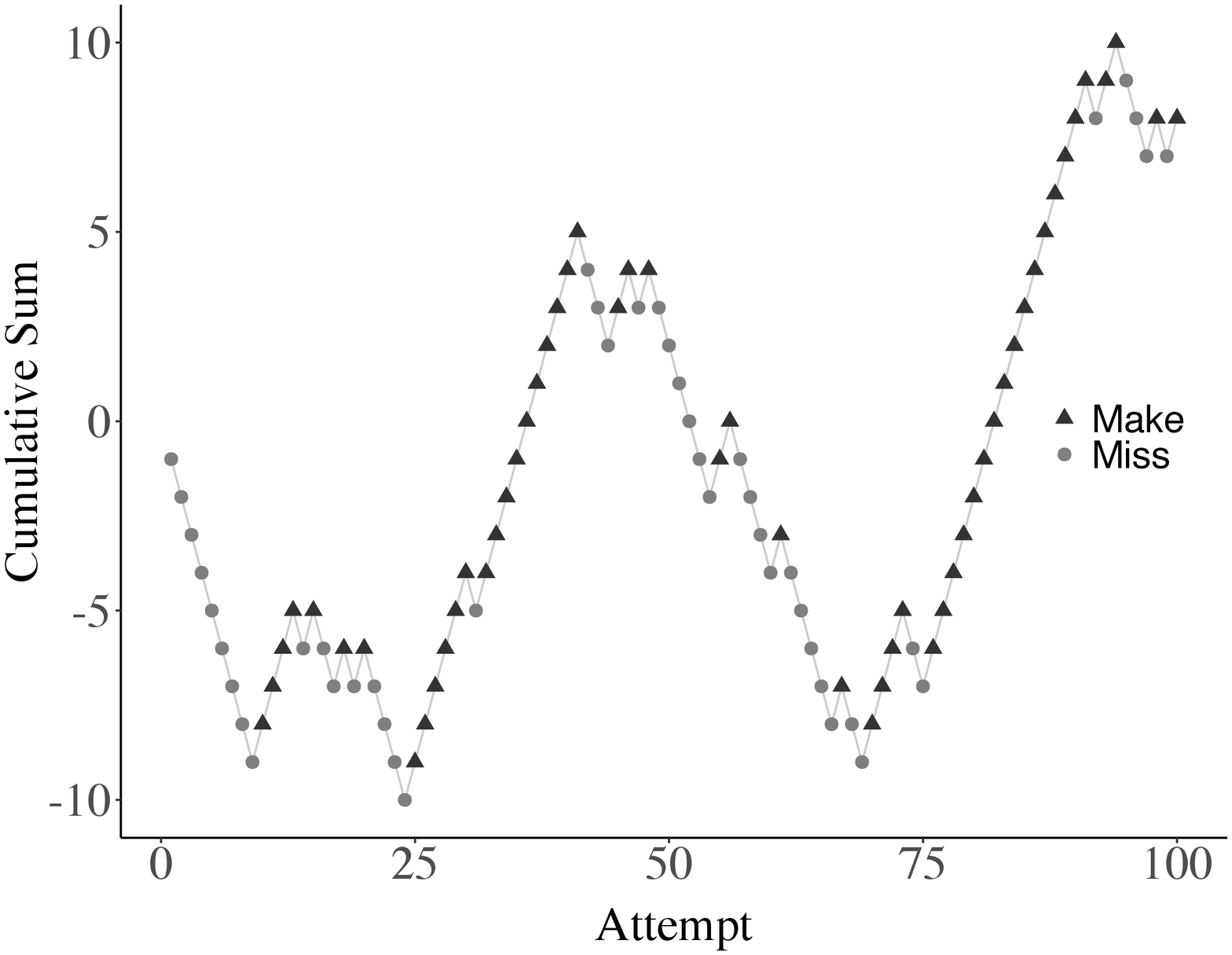}} & \textit{\includegraphics[scale=0.355]{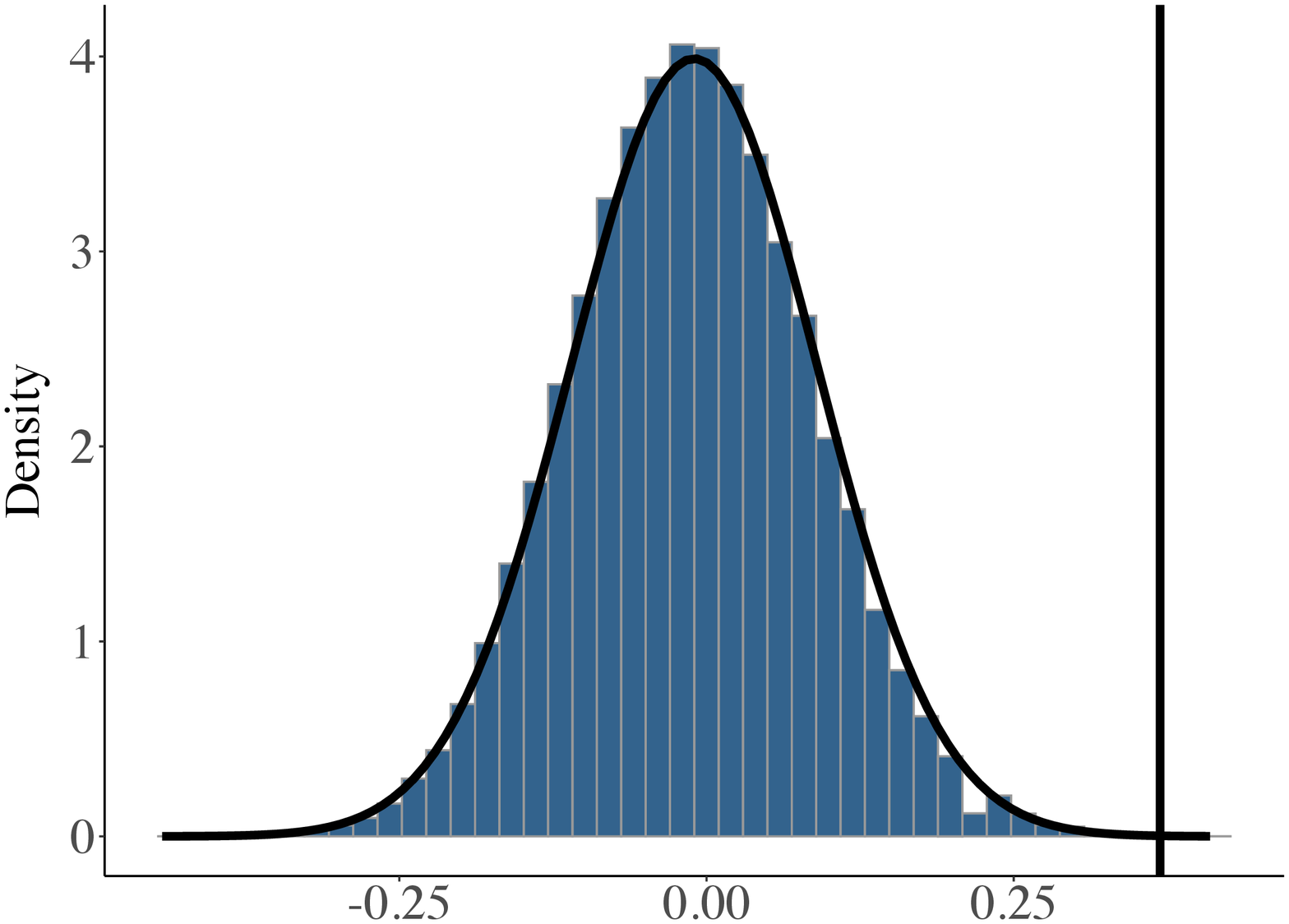}}\tabularnewline
\end{tabular}
\par\end{centering}
\medskip{}

{\footnotesize{}Notes: Panel A displays the cumulative sum of the
sequence of makes and misses for Shooter $109$. Made baskets are
coded as a $1$ and displayed with a black triangle and missed baskets
are coded as a $-1$ and displayed as a grey circle. Panel B displays
a density histogram of $\hat{D}_{n,1}\left(\mathbf{X}_{i}\right)$
computed for $100,000$ permutations of shooter $109$'s observed
shooting sequence. The observed value of $\hat{D}_{n,1}\left(\mathbf{X}_{i}\right)$
is displayed with a vertical black line. The density histogram is
superimposed with $N\left(\beta_{D}^{n,1}\left(p_{i}\right),n^{-1}\sigma_{D}^{2}\left(\hat{p}_{n,i},1\right)\right)$
in black, which is the asymptotic approximation for the permutation
distribution of $\hat{D}_{n,1}\left(\mathbf{X}_{i}\right)$ derived
in Theorem \ref{thm: Convergence_under_H0}, where $\sigma_{D}^{2}\left(\hat{p}_{n,i},1\right)$
is given in the statement of Theorem \ref{thm: Convergence_under_H0},
shifted by the small-sample bias $\beta_{D}^{n,1}\left(p_{i}\right)$.
By Theorem 4 of MS, for $k=1$, we have that $\beta_{D}^{n,1}\left(p_{i}\right)=-1/\left(n-1\right)$.}{\footnotesize\par}
\end{figure}

The rejection of $H_{0}^{i}$ for Shooter 109 in the GVT experiment
for most test statistics, robust to standard multiple hypothesis testing
corrections, is strong evidence that some basketball players exhibit
streaky shooting some of the time. The substantial extent to which
Shooter 109 deviates from randomness is emphasized by Panel A of Figure
\ref{fig:109}, which plots his sequence of makes and misses. Shooter
109 begins by missing 9 shots in a row. Shortly thereafter, he makes
16 out of 17 shots, followed by a sequence where he misses 15 out
of 18 shots and a sequence where he makes 16 shots in a row.

It is unlikely that a random Bernoulli sequence would generate this
pattern, even among $s=26$ sequences.\footnote{GVT observe that the rejection of the individual hypothesis $H_{0}^{i}$
of Shooter 109 is significant, but neither GVT nor MS consider the
multiple testing problem.} Panel B of Figure \ref{fig:109} plots the permutation distribution
of $\hat{D}_{n,1}\left(\mathbf{X}_{i}\right)$ for Shooter 109's shooting
sequence, denoting the observed value of $\hat{D}_{n,1}\left(\mathbf{X}_{i}\right)$
with a vertical black line and our asymptotic approximation to this
distribution with a black curve. The $p$-value of the individual
permutation test using $\hat{D}_{n,1}\left(\mathbf{X}_{i}\right)$
for Shooter 109 is given by the proportion of permutations with recomputed
statistics that are to the right of the observed value; this $p$-value
is equal to $0.0001$.

\begin{figure}[p]
\caption{\label{fig:perm strat}Stratified Permutation Tests of $H_{0}$ using
$\bar{D}_{k}\left(\mathbf{X}\right)$ and $\bar{P}_{k}\left(\mathbf{X}\right)$}

\bigskip{}

\begin{centering}
\begin{tabular}{c}
\textit{\includegraphics[scale=0.33]{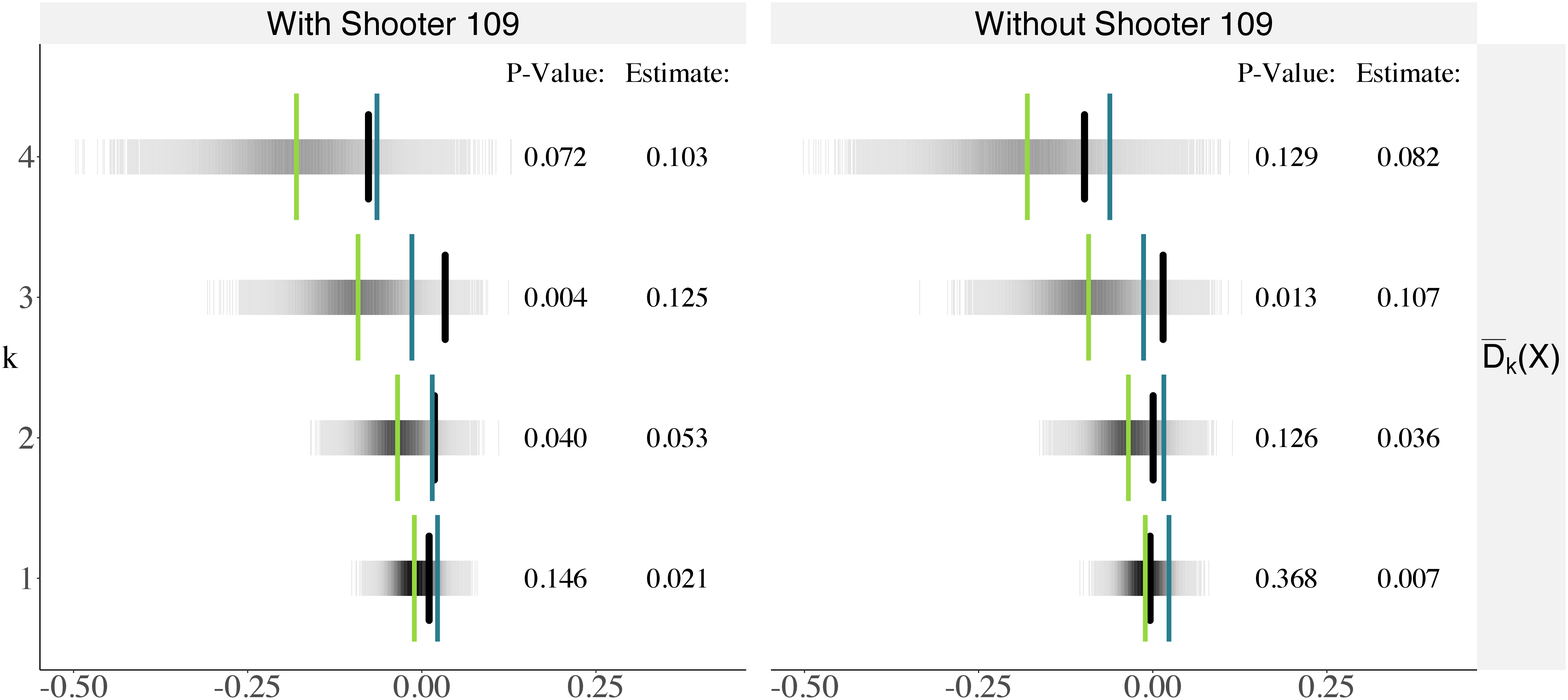}}\tabularnewline
\multicolumn{1}{c}{\textit{\includegraphics[scale=0.33]{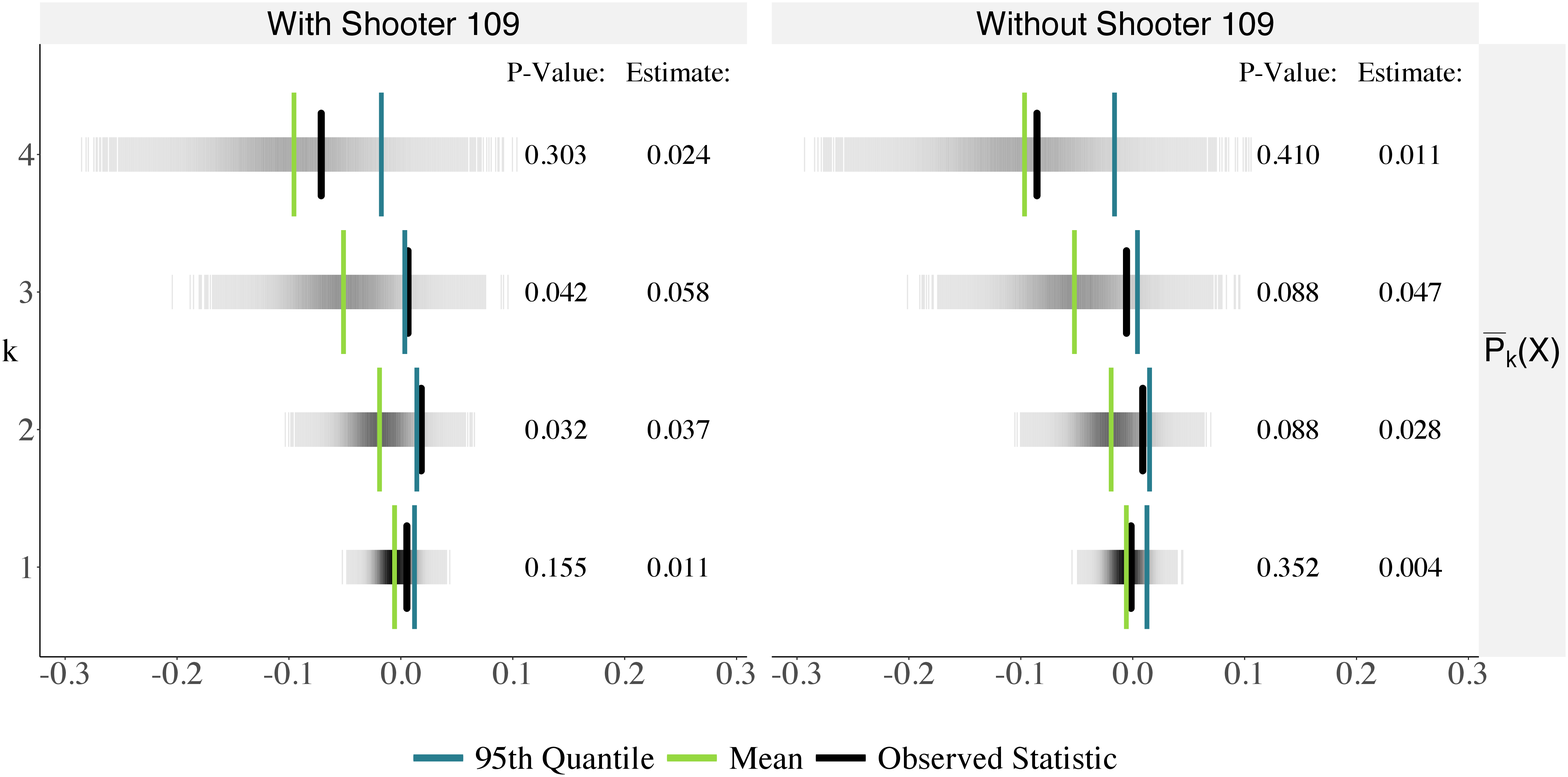}}}\tabularnewline
\end{tabular}
\par\end{centering}
\bigskip{}

{\footnotesize{}Notes: Figure displays the observed values of $\bar{D}_{k}\left(\mathbf{X}\right)$
and $\bar{P}_{k}\left(\mathbf{X}\right)$ overlaid onto the stratified
permutation distribution of $\bar{D}_{k}\left(\mathbf{X}\right)$
and $\bar{P}_{k}\left(\mathbf{X}\right)$ under $H_{0}$ for each
$k$ in $1,\ldots,4$. The observed values of $\bar{D}_{k}\left(\mathbf{X}\right)$
and $\bar{P}_{k}\left(\mathbf{X}\right)$ are indicated by vertical
line segments with squared ends. The estimated $95^{\text{th}}$ quantile
and mean of the permutation distributions under $H_{0}$ are denoted
by vertical line segments with squared ends, respectively. We estimate
the permutation distribution of $\bar{D}_{k}\left(\mathbf{X}\right)$
and $\bar{P}_{k}\left(\mathbf{X}\right)$ under $H_{0}^{i}$ by permuting
each of the $\mathbf{X}_{i}$'s 100,000 times separately and recomputing
$\bar{D}_{k}\left(\mathbf{X}\right)$ and $\bar{P}_{k}\left(\mathbf{X}\right)$
for each permuted collection of sequences. The estimates of the permutation
distribution are displayed in horizontal white to black gradients,
shaded by the proportion of permutations whose recomputed values of
$\bar{D}_{k}\left(\mathbf{X}\right)$ or $\bar{P}_{k}\left(\mathbf{X}\right)$
that lie in a fine partition of the x-axis. The $p$-values of the
stratified permutation test using $\bar{D}_{k}\left(\mathbf{X}\right)$
or $\bar{P}_{k}\left(\mathbf{X}\right)$ are reported to the right
of each distribution for each $k$. The difference between the observed
values of $\bar{D}_{k}\left(\mathbf{X}\right)$ and $\bar{P}_{k}\left(\mathbf{X}\right)$
and the means of the permutation distributions, denoted by $\bar{\tilde{P}}_{k}\left(\mathbf{X}\right)$
and $\bar{\tilde{D}}_{k}\left(\mathbf{X}\right)$ and defined in (\ref{eq: corrected averages}),
are displayed to the right of the $p$-values.}{\footnotesize\par}
\end{figure}

In fact, any evidence of positive dependence in the GVT data appears
to be confined to Shooter 109. Figure \ref{fig:perm strat} overlays
the realized values of $\bar{D}_{k}\left(\mathbf{X}\right)$ and $\bar{P}_{k}\left(\mathbf{X}\right)$
from the GVT experiment on their stratified permutation distributions,
displayed with horizontal black to white gradients, with and without
the inclusion of Shooter 109. The $95^{\text{th}}$ quantiles of these
distributions are denoted by lines with squared ends. The observed
statistics are denoted with vertical lines with rounded ends. The
$p$-values of the stratified permutation tests are displayed to the
right of the corresponding permutation distributions. When Shooter
109 is removed from the sample, only the joint test using $\bar{D}_{k}\left(\mathbf{X}\right)$
for $k$ equal to $3$ is significant at the $5\%$ level.\footnote{In Online Appendix G.4, we implement a similar exercise for joint
tests that use three different statistics as well as two joint testing
methods that combine the results of several tests. The finding that
joint tests are no longer significant after the removal of Shooter
109 is robust to these different choices of test statistics.}

These results are broadly consistent with the evidence from \citet{miller2018cold}
and \citet{miller2019fallacy}. \citet{miller2018cold} report that
$p$-values from the stratified permutation tests of $H_{0}$ using
$\bar{P}_{k}\left(\mathbf{X}\right)$ for $k$ equal to $3$ are equal
to 0.008 and 0.341 using the data from their experiment and from the
\citet{Jagacinski1979} experiment, respectively, but do not report
these results for other values of $k$ or for tests using $\bar{D}_{k}\left(\mathbf{X}\right)$.
Likewise, they highlight one player from their experiment and one
player from the \citet{Jagacinski1979} experiment that are uniquely
streaky. The $p$-values of the permutation tests of $H_{0}^{i}$
using $\hat{P}_{n,k}(\mathbf{X}_{i})-\hat{p}_{n,i}$ with $k$ equal
to $3$ are equal to 0.003 and 0.0001, respectively. Likewise, in
the analysis of the NBA Three Point Shooting contest in \citet{miller2019fallacy},
the stratified permutation test of $H_{0}$ using $\bar{P}_{k}\left(\mathbf{X}\right)$
for $k$ equal to $3$ has a $p$-value less than $0.001$. While
they do not report $p$-values of individual tests, correct for multiplicity,
or report the results of the individual tests using $\hat{P}_{n,k}(\mathbf{X}_{i})-\hat{p}_{n,i}$
or $\hat{D}_{n,k}(\mathbf{X}_{i})$, there is one player with an abnormally
significant rejection of $H_{0}^{i}$ using a more complicated test
statistic.

\subsection{Which Streaky Alternatives Can Be Detected?\label{subsec: power}}

A tempting conclusion from the analysis presented in the previous
subsection might be that streakiness in basketball shooting is confined
to a small number of shooters -- that is, there are a small number
of shooters with very hot hands. However, we argue that (i) the deviation
from randomness exhibited by Shooter 109 is unlikely to be indicative
of what could realistically be expected from even a small proportion
of basketball players and that (ii) the existing controlled shooting
experiments do not have sufficient power to detect what would be realistic
alternatives.

In support of the former point, Table \ref{tab: Shooter 109 Estimates}
displays the statistics $\tilde{P}_{n,k}(\mathbf{X}_{i})$ and $\tilde{D}_{n,k}(\mathbf{X}_{i})$,
defined in (\ref{eq: corrected}), computed on Shooter 109's shooting
sequence. We caution that these statistics are only bias-corrected
under the null hypothesis. Nevertheless, taken as estimates of $\theta_{P}^{k}\left(\mathbb{P}_{i}\right)$
and $\theta_{D}^{k}\left(\mathbb{P}_{i}\right)$, they correspond
to massive deviations from randomness. 
\begin{table}[t]
\begin{centering}
\begin{tabular}{cccc}
\toprule 
k &  & $\tilde{P}_{n,k}(\mathbf{X}_{i})$ & $\tilde{D}_{n,k}(\mathbf{X}_{i})$\tabularnewline
\midrule
\midrule 
1 &  & 0.182 & 0.379\tabularnewline
\midrule
2 &  & 0.263 & 0.487\tabularnewline
\midrule
3 &  & 0.324 & 0.561\tabularnewline
\midrule
4 &  & 0.330 & 0.593\tabularnewline
\bottomrule
\end{tabular}
\par\end{centering}
{\small{}\medskip{}
}{\small\par}

\caption{\label{tab: Shooter 109 Estimates}Estimates of $\theta_{P}^{k}\left(\mathbb{P}_{i}\right)$
and $\theta_{D}^{k}\left(\mathbb{P}_{i}\right)$ for Shooter 109}

\medskip{}

{\footnotesize{}Notes: Table displays the statistics $\tilde{P}_{n,k}(\mathbf{X}_{i})$
and $\tilde{D}_{n,k}(\mathbf{X}_{i})$, defined in (\ref{eq: corrected}),
computed using data from Shooter 109 from the GVT controlled shooting
experiment.}{\footnotesize\par}
\end{table}

\begin{figure}[tp]
\caption{\label{fig:NBA shooting}Distribution of Field Goal Shooting Percentage
in the 2018-2019 NBA Season}

\medskip{}

\begin{centering}
\begin{tabular}{cc}
\textit{\small{}Panel A: Empirical Distribution Function} & \textit{\small{}Panel B: Histogram}\tabularnewline
\textit{\small{}\includegraphics[scale=0.3]{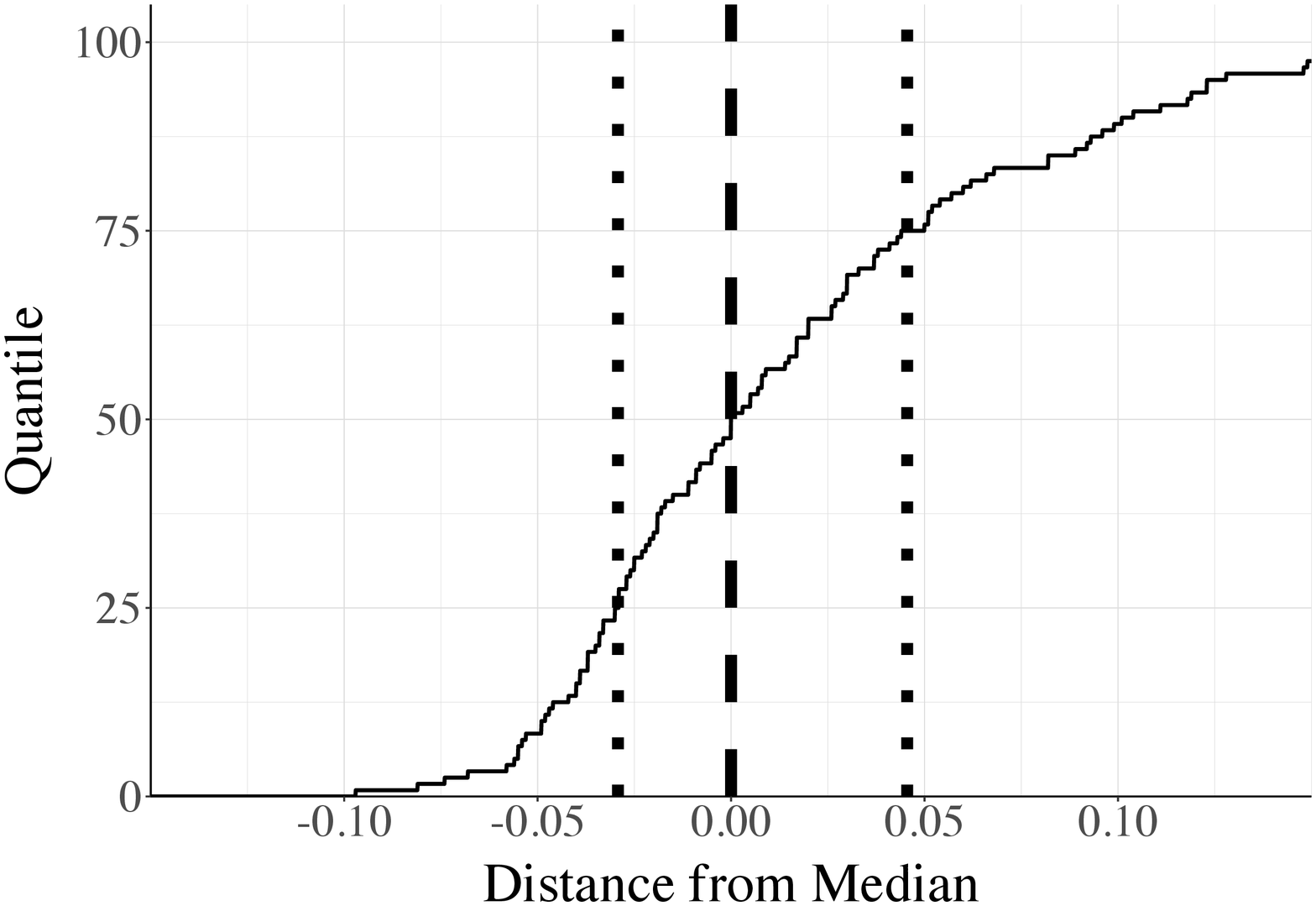}} & \includegraphics[scale=0.3]{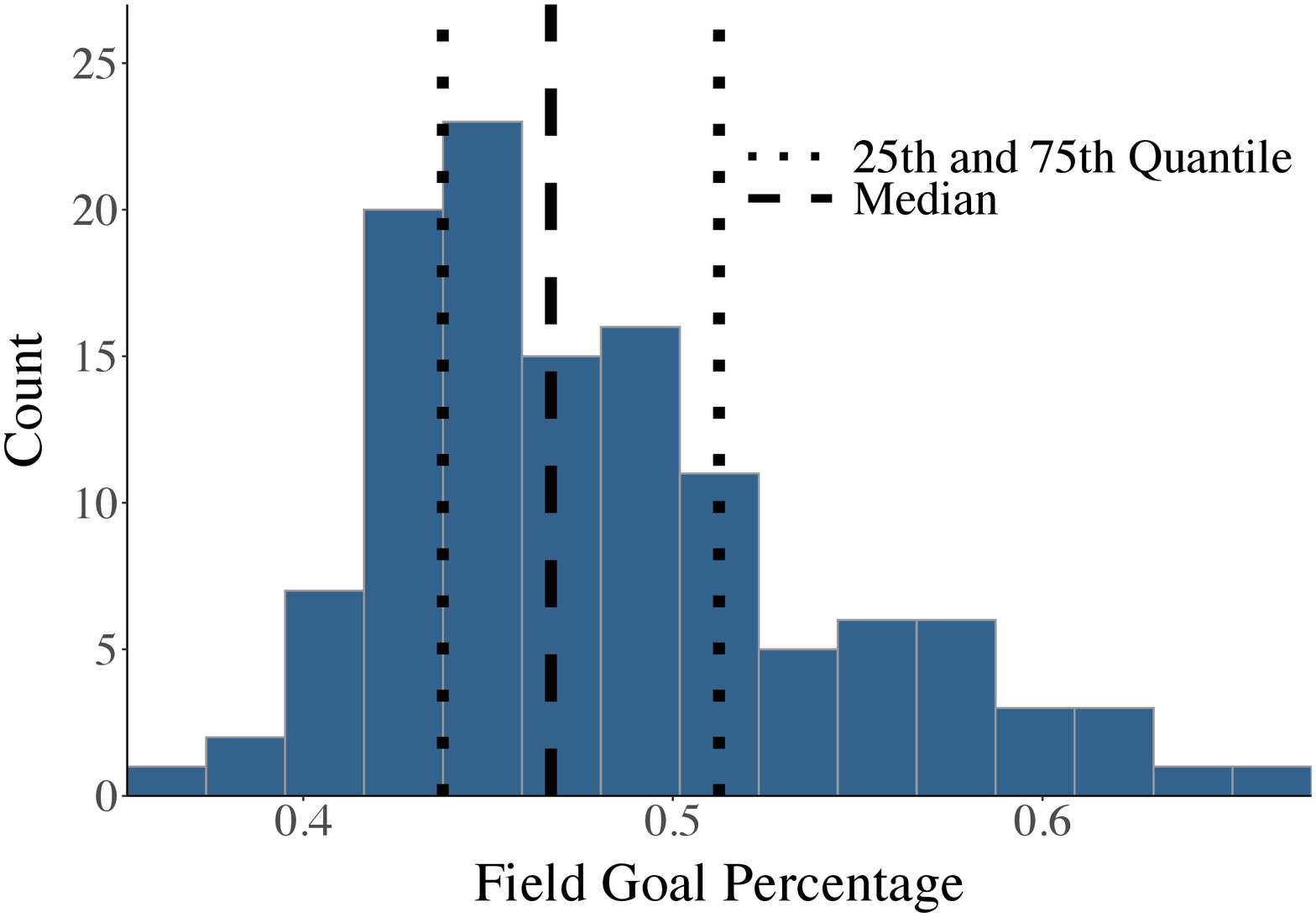}\tabularnewline
\end{tabular}
\par\end{centering}
\medskip{}

{\footnotesize{}Notes: Figure displays the distribution of the field
goal shooting percentages of NBA players in the 2018--2019 regular
season. Players shooting fewer than 300 field goals are omitted when
displaying the distributions. The left panel displays a truncated
empirical cumulative distribution and the right panel displays a histogram
of the shooting percentages. To parallel the Markov streaky alternatives
specified in Section \ref{subsec:A-Markov-Model}, the x-axis of the
truncated cumulative distribution is transformed such that the median
is displayed as 0, and $\epsilon$ corresponds to the difference,
in terms of shooting percentage, between the x-axis position and the
median. The vertical dashed lines give the medians of the distributions.
The vertical dotted lines give the $25^{\text{th}}$ and $75^{\text{th}}$
quantiles of the distributions.}{\footnotesize\par}
\end{figure}

For perspective, Figure \ref{fig:NBA shooting} displays a histogram
and an empirical distribution function of the field goal percentages
of NBA players in the 2018--2019 regular season.\footnote{The data were downloaded from \citet{basketball_reference}. The field
goal sample includes players who have attempted more than 300 field
goals. Field goals are shots taken in any context of a live basketball
game, other than free throws.} The x-axis of the empirical distribution function plot has been relabelled
such that the median of the distribution is displayed as $0$. The
statistic $\tilde{D}_{n,k}(\mathbf{X}_{i})$ with $k$ equals one
for Shooter 109 is equal to 0.379. Taken as an estimate of $\theta_{D}^{k}\left(\mathbb{P}_{i}\right)$,
this corresponds to varying between shooting at a rate similar to
the best or worst shooter in the NBA, depending on whether a shooter
made or missed their previous shot.

Now, perhaps more realistically, suppose that the marginal shooting
percentage for all shooters is 50\%. For half of these shooters, shooting
percentage increases and decreases by half of the interquartile range
of the distribution of field goal percentages of NBA players after
making their previous $m$ shots or missing their previous $m$ shots,
respectively. The other half of shooters remain 50\% shooters at all
times. This is an instance of Markov chain streaky alternative studied
in Section \ref{subsec:A-Markov-Model}, parameterized as $\epsilon=0.038$
and $\zeta=0.5$ for a given value of $m$.\footnote{A similar parameterization is obtained if we consider half the distance
between the top and bottom terciles of the NBA free throw distribution.
In fact, in this case, $\epsilon=0.0384$. However, the distribution
of free throw percentages has a larger median (0.806) and a long left
tail. The data were downloaded from \citet{basketball_reference}.
The free throw sample includes players who have attempted more than
125 free throws.}

We argue that this parameterization is a conservative upper bound
on the set of deviations from randomness consistent with the variation
in NBA shooting percentages.\footnote{Note that, additionally, the power of the joint permutation tests
using $\tilde{D}_{n,k}(\mathbf{X}_{i})$ decreases if marginal shooting
percentages are different than $1/2$ or are sampled from a distribution.} In our choice of $\epsilon$, in effect, we assume that the variation
in shooting percentages within players is less than the variation
in shooting percentages across players. As the proportion of players
in each experiment with large values of $\tilde{P}_{n,k}(\mathbf{X}_{i})$
and $\tilde{D}_{n,k}(\mathbf{X}_{i})$ is small, imposing $\zeta=0.5$
is likely to be very conservative.\footnote{See Online Appendix Figures 7 and 8.}
We consider two relaxations of this upper bound, which in our judgment
do not seem less reasonable: reducing the proportion of streaky individuals
to 25\% ($\zeta=0.25)$ and assuming that streaky individuals increase
and decrease their shooting percentages by half the distance between
the 66$^{\text{th}}$ and 33$^{\text{rd}}$ quantiles of the NBA field
goal percentage distribution after making or missing $m$ shots $(\epsilon=0.024)$.

We consider $m=3$ as the benchmark parameterization based on the
emphasis in GVT and MS, although parameterizations with $m=1$ give
a more conservative upper bound on power. Specifically, GVT describe
streaks of three makes (misses) as ``hot'' (``cold'') periods
and emphasize statistics with $k=3$. Likewise, \citet{miller2018cold}
and \citet{miller2019fallacy} denote streaks of three makes (misses)
as ``hot'' (``cold'') streaks. \citet{miller2018cold} argue that
they are interested in detecting alternatives with $m=3$, citing
literature in psychology indicating that people perceive streaks to
begin at three \citep{carlson2007rule} and only implementing statistics
with $k$ equal to 3.

\begin{figure}[p]
\caption{\label{fig: approximate}Power of Joint Tests of $H_{0}$ for Four
Controlled Basketball Shooting Experiments}

\medskip{}

\begin{centering}
\begin{tabular}{c}
\multicolumn{1}{c}{\includegraphics[scale=0.4]{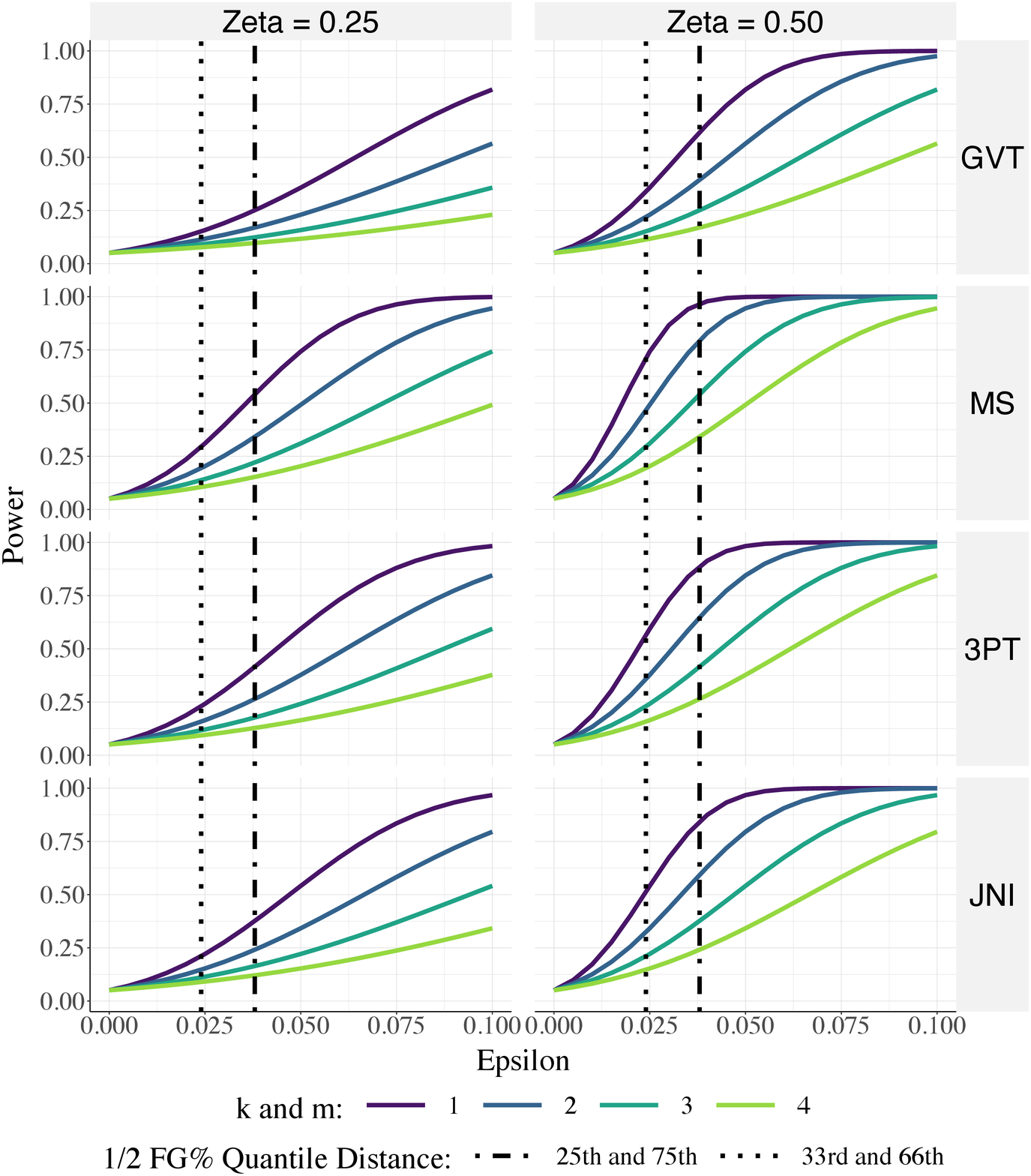}}\tabularnewline
\end{tabular}
\par\end{centering}
\medskip{}

{\footnotesize{}Notes: Figure displays asymptotic approximations to
the power of the stratified permutation tests of the joint null hypothesis
$H_{0}$ using the test statistic $\bar{D}_{k}\left(\mathbf{X}\right)$
against the Markov chain streaky alternative specified in Section
\ref{subsec:A-Markov-Model} over a grid of $\epsilon$, for $\zeta$
equal to $0.25$ or $0.5$, for $m=k$ for $k$ in $1,\ldots,4$,
and for $n$ and $s$ at the values of the GVT, \citet{miller2018cold}
(MS), \citet{Jagacinski1979} (JNI), and \citet{miller2019fallacy}
(3PT) controlled basketball shooting experiments. An expression for
this asymptotic approximation is given in Online Appendix Corollary
H.3, with $h=\epsilon\sqrt{ns}$. The vertical dotted and dot-dashed
black lines denote the $\epsilon=0.024$ and $\epsilon=0.038$, which
are consistent with half the distance between the 33$^{\text{rd}}$
and the $66$$^{\text{th}}$ quantiles and the 25$^{\text{th}}$ and
the $75$$^{\text{th}}$ quantiles of the distribution of NBA field
goal percentages, respectively. Note that in the setting of \citet{miller2019fallacy},
the NBA Three Point Shooting contest, participants take different
numbers of shots. For this example we replace $ns$ with $s$ times
the average number of shots taken, which gives an approximation to
the power of the stratified permutation test using the sample-size
weighted average of $\hat{D}_{n,k}\left(\mathbf{X}_{i}\right)$ across
individuals. }{\footnotesize\par}
\end{figure}

Figure \ref{fig: approximate} displays our asymptotic approximation
to the power of the stratified permutation test of $H_{0}$ using
the test statistic $\bar{D}_{k}\left(\mathbf{X}\right)$ against the
Markov chain streaky alternative studied in Section \ref{subsec:A-Markov-Model}
over a grid of $\epsilon$, for $\zeta$ equal to $0.25$ and $0.5$,
$m=k$ for $k$ in $1,\ldots,4$, and for $n$ and $s$ equal to their
values in the GVT, \citet{miller2018cold}, \citet{Jagacinski1979},
and \citet{miller2019fallacy} controlled basketball shooting experiments.\footnote{Note that in the setting of \citet{miller2019fallacy} -- the NBA
Three Point Shooting contest -- participants take different numbers
of shots. For this example we replace $ns$ with $s$ times the average
number of shots taken, obtained from Table 1 of \citet{miller2019fallacy}
, which gives an approximation to the power of the stratified permutation
test using the sample-size weighted average of $\hat{D}_{n,k}\left(\mathbf{X}_{i}\right)$
across individuals.} The vertical black lines denote $\epsilon=0.024$ and $\epsilon=0.038$.

All four experiments lack adequate power against conservative parameterizations
of the Markov chain streaky alternative specified above.\footnote{It follows that multiple tests of the individual hypotheses are even
less powerful. Indeed, as any multiple hypothesis testing method that
controls the FWER must be constructed with the closure method \citep{romano2011consonance},
and under the closure method the individual hypothesis $H_{0}^{i}$
is rejected if all joint tests of subsets of $\left\{ H_{0}^{i}:i\in1,\dots,s\right\} $
containing $H_{0}^{i}$ are rejected, then even the rejection of the
joint null hypothesis $H_{0}$ is not sufficient to obtain any rejections
of the individual hypotheses $H_{0}^{i}$.} No experiment has adequate power for $\zeta=0.25$ or for $\epsilon=0.024$
for any value of $m$. No experiment has adequate power for $\epsilon=0.038$
and $\zeta=0.5$ for $m=3$. For $\epsilon=0.038$ and for $\zeta=0.5$,
\citet{miller2018cold} and \citet{miller2019fallacy} have reasonable
power for $m=1$ and \citet{miller2018cold} has reasonable power
for $m=2$.

\subsection{How Much Positive Serial Dependence is There in Basketball Shooting?}

An answer to the second question -- how widespread and substantial
is dependence in basketball shooting --- can be assessed with estimates
of the individual parameters $\theta_{P}^{k}\left(\mathbb{P}_{i}\right)$
and $\theta_{D}^{k}\left(\mathbb{P}_{i}\right)$ and the average parameters
$\bar{\theta}_{P}^{k}\left(\mathbb{P}\right)$ and $\bar{\theta}_{D}^{k}\left(\mathbb{P}\right)$.
However, measurement of the magnitude of the average streakiness of
basketball shooting can only be distinguished from zero if reasonable
deviations from randomness can be detected, and as we have argued
in the previous subsection, this is not the case for the existing
controlled shooting experiments.

Returning to Figure \ref{fig:perm strat} and the GVT shooting experiment,
the statistics $\bar{\tilde{P}}_{k}\left(\mathbf{X}\right)$ and $\bar{\tilde{D}}_{k}\left(\mathbf{X}\right)$,
defined in (\ref{eq: corrected}), are given by the difference between
the observed statistics and the means of their permutation distributions
and are displayed under ``Estimate'' on the far right hand side
of each panel. These statistics are unbiased for $\bar{\theta}_{P}^{k}\left(\mathbb{P}\right)$
and $\bar{\theta}_{D}^{k}\left(\mathbb{P}\right)$ under $H_{0}$.
These estimates are large for $k$ equal to $2$ and $3$, and particularly
for $\bar{\theta}_{D}^{k}\left(\mathbb{P}\right)$ with $k$ equal
to 3. However, with the exception of $\bar{\theta}_{D}^{k}\left(\mathbb{P}\right)$
for $k=3$, the permutation tests no longer reject at the $5\%$ level
after the removal of Shooter 109.

Recall that in Theorem \ref{thm: exact permutation}, we show that
permutation tests are the only tests with exact type 1 error control.
This implies that if a stratified permutation test of $H_{0}$ using
the test statistic $K_{n,s}$ does not reject the null hypothesis,
then any lower confidence bound constructed with test inversion using
the test statistic $K_{n,s}$ that obtains exact $95\%$ coverage
will be below zero. Thus, with the exception of $\bar{\theta}_{D}^{k}\left(\mathbb{P}\right)$
for $k=3$, any lower confidence bound for the parameters $\bar{\theta}_{P}^{k}\left(\mathbb{P}\right)$
and $\bar{\theta}_{D}^{k}\left(\mathbb{P}\right)$ constructed with
statistics $\bar{\tilde{P}}_{k}\left(\mathbf{X}\right)$ and $\bar{\tilde{D}}_{k}\left(\mathbf{X}\right)$
using test inversion that obtains exact $95\%$ coverage will be below
zero. The analogous lower confidence bound for $\bar{\theta}_{D}^{k}\left(\mathbb{P}\right)$
for $k=3$ will be close to zero.

In practice, for experiments with significantly larger samples, we
advocate for the application of general bootstrap methods for stationary
time series (see \citet{lahiri2013resampling}), such as the moving
blocks bootstrap \citep{liu1992moving,kunsch1989jackknife}, the stationary
bootstrap \citep{politis1994stationary}, or subsampling \citep{politis1999subsampling}.\footnote{Additionally, simultaneous confidence regions for the individual parameters
$\theta_{P}^{k}\left(\mathbb{P}_{i}\right)$ and $\theta_{D}^{k}\left(\mathbb{P}_{i}\right)$
can be constructed using a \v{S}id\'{a}k correction for multiplicity.}

To conclude, existing randomized shooting experiments are insufficiently
powered to detect deviations from randomness that we argue would be
consistent with a realistic parameterization of positive dependence
in basketball shooting. These experiments are therefore unable to
provide an informative estimate of the mean or dispersion of the serial
dependence in basketball shooting. This conclusion could be challenged
by a strong and robust rejection of $H_{0}$, but the rejection of
$H_{0}$, at least in the case of the GVT experiment, is sensitive
to inclusion of an outlier. This result cuts both ways. The data are
insufficient to make strong statements about the magnitude of positive
dependence in basketball shooting, either small or substantial.

\subsection{Do People Overestimate Positive Serial Dependence?}

If we had an informative estimate of the positive serial dependence
of an average shooter, a comparison with evidence on expectations
of serial dependence in basketball shooting would provide a direct
test of the hot hand fallacy. Specifically, we advocate for a test
of the null hypothesis that $\bar{\theta}_{P}^{k}\left(\mathbb{P}\right)$
and $\bar{\theta}_{D}^{k}\left(\mathbb{P}\right)$ are equal to an
audience's expectations of $\bar{\theta}_{P}^{k}\left(\mathbb{P}\right)$
and $\bar{\theta}_{D}^{k}\left(\mathbb{P}\right)$ against the alternative
that the audience's expectations are larger. We find that the available
evidence on expectations of streakiness in basketball shooting suffers
either from prohibitive methodological flaws or is not directly comparable
to estimates of $\bar{\theta}_{P}^{k}\left(\mathbb{P}\right)$ and
$\bar{\theta}_{D}^{k}\left(\mathbb{P}\right)$.

GVT measure beliefs with two methods. First, they implement a survey
of one hundred basketball fans from Cornell and Stanford. The fans
were asked to consider a hypothetical basketball player who makes
$50\%$ of their shots. The average expected field goal percentages
for this player after having just made and missed a shot were $61\%$
and $42\%$, respectively. Similarly, when asked to consider a hypothetical
player who makes $70\%$ of shots from the free throw line, fans expected
that the average free throw percentages for second free throws after
having made and missed the first were $74\%$ or $66\%$, respectively.

Taken at face value, the surveys can be interpreted as eliciting expectations
of $\theta_{D}^{k}\left(\mathbb{P}_{i}\right)$ when $k=1$ and indicating
that these expectations are approximately $0.1$ and $0.04$, respectively.
However, there are severe methodological limitations to the GVT survey.
First, there is considerable evidence that surveys eliciting beliefs
about hypothetical events can be prone to substantial bias \citep{harrison2008experimental}.
Second, the results may be biased by framing \citep{tversky1981framing};
that is, the language of the survey questions may be suggestive of
positive serial dependence.

Second, GVT attempt to infer beliefs from observations of incentivized
decisions. In their controlled shooting experiment, prior to each
shot, each shooter and an observer choose whether to bet ``high''
or ``low.'' If an individual bets high (low) and makes the shot
they win $5$ ($2$) cents. If the individual bets high (low) and
misses the shot they lose $4$ ($1$) cents.\footnote{These data are not publicly available.}
\citet{miller2017visible} find that the average correlation between
the bets and the shot outcomes is $0.07$, that the increase in the
probability of predicting a make after a make is 0.077, and that these
estimates are significantly different from zero.

Unfortunately, these estimates do not pin down an estimate of $\bar{\theta}_{P}^{k}\left(\mathbb{P}\right)$
or $\bar{\theta}_{D}^{k}\left(\mathbb{P}\right)$. Assume that individuals
only bet on a make if they believe that there is greater than a 50\%
chance of a make. In this case, if we observe infinite shots, the
proportion of shots in which an individual predicts a make is equal
to the proportion of shots in which the individual expects that the
probability of a make is greater than 50\%. This proportion is not
in general equal to the individual's average expectation of the probability
of a make.

\citet{miller2018cold} implement a survey of the participants in
their experiment, asking the basketball players to rate how likely
their teammates are to make a shot following a sequence of three makes
on an arbitrary scale from -3 to 3. Again, this does not identify
an estimate of the player's expectations of the serial dependence
in basketball shooting.

In our review of the literature, we are unable to find estimates of
beliefs in serial dependence in basketball shooting that directly
translate to estimates of people's expectations of $\bar{\theta}_{P}^{k}\left(\mathbb{P}\right)$
or $\bar{\theta}_{D}^{k}\left(\mathbb{P}\right)$. \citet{rao2009experts},
\citet{bocskocsky2014hot}, and \citet{lantis2019hot} explore shot
selection and defensive pressure in NBA games, and find that players
behave as if they believe that the probability of a make is higher
after a streak of makes than after a streak of misses. Again, these
studies do not provide an estimate of beliefs directly comparable
to estimates of $\bar{\theta}_{P}^{k}\left(\mathbb{P}\right)$ or
$\bar{\theta}_{D}^{k}\left(\mathbb{P}\right)$.

Future studies should estimate expectations of $\bar{\theta}_{P}^{k}\left(\mathbb{P}\right)$
or $\bar{\theta}_{D}^{k}\left(\mathbb{P}\right)$ that are directly
comparable to measurements of $\bar{\theta}_{P}^{k}\left(\mathbb{P}\right)$
or $\bar{\theta}_{D}^{k}\left(\mathbb{P}\right)$. \citet{manski2004measuring}
advocates for surveys of probabilistic expectations in non-hypothetical
settings. Data from surveys of this form have been valuable in informing
behavioral models of expectation formation in financial markets \citep{greenwood2014expectations,barberis2015x}.
We support a design in which an observer of a shooter in a controlled
shooting experiment is asked to record their expectation of the probability
that the shooter makes their next shot prior to each shot. If the
shot is made, the observer is rewarded for submitting large probabilities
and punished for submitting small probabilities. If the shot is missed,
the converse is true. This design would not suffer from the framing
bias of the GVT survey and would provide a direct estimate of beliefs
in $\bar{\theta}_{P}^{k}\left(\mathbb{P}_{i}\right)$ and $\bar{\theta}_{D}^{k}\left(\mathbb{P}_{i}\right)$.
Nevertheless, it is important to ensure that these measurements of
beliefs are made on a sample large enough to ensure adequate precision
for an informative comparison to measurements of $\bar{\theta}_{P}^{k}\left(\mathbb{P}_{i}\right)$
and $\bar{\theta}_{D}^{k}\left(\mathbb{P}_{i}\right)$.

\section{Conclusion\label{sec:Conclusion}}

The purpose of this paper is to clarify and quantify the uncertainty
in the empirical support for the tendency to perceive streaks as overly
representative of positive dependence -- the hot hand fallacy. Following
\citet*{gilovich1985hot}, the results of a class of tests of randomness
implemented on data from a basketball shooting experiment have provided
central empirical support for textbook models of misperception of
randomness. The results and conclusions of these tests were called
into question by \citet{miller2018surprised}, who observe that there
is a substantial small sample bias in the test statistics that had
been applied. We evaluate the implications, limitations, and interpretation
of these tests by establishing their validity, approximating their
power, and re-evaluating their application to four controlled basketball
shooting experiments.

Our theoretical and simulation analyses show that the tests considered
are insufficiently powered to detect effect sizes consistent with
the observed variation in NBA shooting percentages with high probability.
Substantially larger data sets are required for informative estimates
of the streakiness in basketball shooting. We are able to reject i.i.d.\
shooting consistently for only one participant in the \citet*{gilovich1985hot}
shooting experiment. This rejection is robust to standard multiple
testing corrections, providing strong evidence that basketball shooting
is not perfectly random. However, evidence against randomness in that
experiment is limited to this player.

Future research should directly test the accuracy of people's predictions
of streakiness in stochastic processes and should be implemented in
settings with reasonable power against sensible alternatives. We provide
a mathematical and statistical theory to serve as a foundation for
future analyses with this objective. Our analytic power approximations
significantly reduce the computational burden of power analyses in
the design of these studies. Additionally, we contribute an emphasis
on the differentiation of individual, simultaneous, and joint hypothesis
testing that can more clearly delineate the conclusions and limitations
of inferences on deviations from randomness.

\bigskip{}

\end{spacing}

\paragraph{Data Availability Statement:}

The data and code underlying this article are available on Zenondo as ``Replication
package for: Uncertainty in the Hot Hand Fallacy: Detecting Streaky
Alternatives to Random Bernoulli Sequences'' at \href{http://doi.org/10.5281/zenodo.4563661}{http://doi.org/10.5281/zenodo.4563661}.

\pagebreak{}

\bibliographystyle{apa}
\bibliography{hot_hand}

\pagebreak
\end{document}